\documentclass[twocolumn]{aastex62}

\usepackage{savesym}
\savesymbol{tablenum}
\usepackage{siunitx}
\restoresymbol{SIX}{tablenum}

\usepackage{wasysym}

\DeclareSIUnit\au{au}
\DeclareSIUnit\msun{M_\odot}
\DeclareSIUnit\mearth{M_\oplus}
\DeclareSIUnit\myr{Myr}

\def\rapoH{\tilde{r}_\mathrm{apo}}
\def\rcollH{\tilde{r}_\mathrm{coll}}
\def\rperiH{\tilde{r}_\mathrm{peri}}

\defcitealias{2019ApJEmsenhuberA}{EA19}

\received{2019 April 6}
\revised{2019 July 1}
\accepted{2019 July 3}
\published{2019 August 20}
\submitjournal{The Astrophysical Journal}

\shorttitle{GMCs under External Perturbers}
\shortauthors{Emsenhuber and Asphaug}

\begin{document}

\title{Graze-and-Merge Collisions under External Perturbers}

\correspondingauthor{Alexandre Emsenhuber}
\email{emsenhuber@lpl.arizona.edu}

\author[0000-0002-8811-1914]{Alexandre Emsenhuber}
\affil{Lunar and Planetary Laboratory, University of Arizona \\
1629 E. University Blvd. \\
Tucson, AZ 85721, USA}

\author[0000-0003-1002-2038]{Erik Asphaug}
\affil{Lunar and Planetary Laboratory, University of Arizona \\
1629 E. University Blvd. \\
Tucson, AZ 85721, USA}

\begin{abstract}

Graze-and-merge collisions (GMCs) are common multi-step mergers occurring in low-velocity off-axis impacts between similar sized planetary bodies. The first impact happens at somewhat faster than the mutual escape velocity; for typical impact angles this does not result in immediate accretion, but the smaller body is slowed down so that it loops back around and collides again, ultimately accreting. The scenario changes in the presence of a third major body, i.e. planets accreting around a star, or satellites around a planet. We find that when the loop-back orbit remains inside roughly 1/3 of the Hill radius from the target, then the overall process is not strongly affected. As the loop-back orbit increases in radius, the return velocity and angle of the second collision become increasingly random, with no record of the first collision's orientation. When the loop-back orbit gets to about 3/4 of the Hill radius, the path of smaller body is disturbed up to the point that it will usually escape the target.

\end{abstract}


\section{Introduction}
\label{sec:intro}

Graze-and-merge collisions (GMCs) are multi-step processes commonly occurring in relatively low-velocity collisions between similar-sized bodies, better known as giant impacts. In similar-sized collisions (SSCs) there is usually an incomplete intersection between the smaller body (the projectile, also called the impactor) and the target, even for moderate impact angles \citep{2010ChEGAsphaug} so that the bulk of the projectile is not yet stopped. If the projectile continues faster than the escape velocity it is a hit and run collision (HRC). If the projectile dissipates enough kinetic energy in the collision that the bodies emerge gravitationally bound, and if the system is isolated, then the bodies will collide again after one revolution, typically within tens of hours, leading to the actual merger. When the colliding bodies are nearly equal size, GMCs are actually the most common form of accretion \citep{2012ApJStewart}. It is the second collision in a GMC that produces the most pronounced effects of the overall process, spinning up the merged body and in some cases, launching and torquing material into orbit that makes a protolunar disk.

Since SSCs involve large angular momentum, the last few giant impacts determine the spin of the resulting bodies \citep{1999IcarusAgnor,2007ApJKokubo}. GMCs are also invoked to explain the formation of multiple satellite systems or families in the solar system. The canonical scenario for the Moon formation \citep{1986IcarusBenz,2001NatureCanup,2004IcarusCanup} is a GMC where the projectile is initially captured and performs an orbit with a period of the order of a day. The second part of the collision, the merger, is responsible for ejecting material into orbit, from which the Moon eventually accretes. GMCs are also invoked for the cases of minor planets; for instance with the Haumea system and its dynamical family \citep{2010ApJLeinhardt}. The formation of the Pluto-Charo system \citep{2005ScienceCanup,2011AJCanup} has been modeled as a specific case where following the first encounter, the deformation of Pluto is such that it puts a torque on Charon, modifying its orbit so that no further collision occurs.

Similar scenarios have been invoked in planetary systems. \citet{2013IcarusAsphaug} proposed that one or more GMCs during the accretion of Titan by giant impacts could produce Saturn's middle-sized moons as by-products. But in this case, especially for collisions occurring withing 10-20 planetary radii of Saturn, the timescale between the first and second collision in a GMC can be a substantial fraction of the orbital period around the planet, so that the planet's influence is important. The same situation is true of GMCs involving extra-solar planets orbiting close to their stars.

We have found that modeling the same collisions, accounting for the presence of Saturn or another central body, is challenging because many realisations of the same event are necessary to account for the possible orientations of the three bodies (projectile, target, planet). One possibility is to model the different stages of a GMC separately, using hydrodynamical scheme for the encounters themselves (projectile and target), and using an \textit{N}-body code in between to connect them. While this scheme can improve the dynamics and allow for much more effective exploration of impact parameters, it is necessary to understand if and when the different stages can be separated, i.e. when the encounters phases are barely affected by the presence of other bodies.

All the GMCs discussed here were modeled neglecting the presence of other bodies. In \citet[][hereafter \citetalias{2019ApJEmsenhuberA}]{2019ApJEmsenhuberA}, we studied the dynamical evolution of remnants from HRCs. We found that under the influence of other bodies, weakly unbound runners would collide back on the target, in what we call Hit and run returns (HRRs), in only 60\% of the cases. A corollary to this result is that weakly-bound GMCs would not all fit in the general picture. We expect that some of those would end up being HRCs.

This has potential implications for formation studies. Current scaling laws, e.g., \citet{2012ApJLeinhardt} treat those a mergers. However, GMCs whose bodies range beyond a significant fraction of the Hill radius might not come back, hence resulting in HRCs and delaying the formation process \citep{2013IcarusChambers}.

GMCs and HRRs are different in nature. If a GMC is not influenced by external effects, then the properties of the second encounter are determined from the first one; the impact velocity and angles are fixed. In HRRs, we found that the velocity is correlated to the end state of the first collision, though some spread occurs due to following close encounters and secular perturbations, while the impact angle follows the expected distribution if relative positions are uniformly distributed in space \citep{1962BookShoemaker}.

Directly modeling GMCs with the presence of an external body, while accounting for all possible orientations is infeasible. A simple model for this has already three parameters: two angles for the relative orientation of the collision with respect to the body, another one for the direction of motion. This is assuming that the mass of the other body as well as its distance are known, otherwise they add to the list.

Instead of performing full hydrodynamical simulations of such cases, we will assume that a GMC can be divided into multiple stages that can be treated independently: the initial encounter, the following orbital evolution and the further encounters. The focus of this work is on the second stage, that is the orbital evolution in between the collisions. If this phase is well described by the orbit of point masses, then it can be modeled by means of inexpensive \textit{N}-body calculations. This allows us to perform many realisations assuming different orientations to derive relationships between orientation, distance, and the expected outcome.

Our aim is to determine when the presence of bodies affects the process of a GMC. We will focus on the effects of the bodies orbiting around a central star (or planet) in a first stage, but we will also cover the case when other planets (or satellites) are present in the same system.

\section{Dynamics of a graze and merge collision}
\label{sec:gmc}

The first encounter of a GMC is similar as in a HRC \citep{2006NatureAsphaug}. It dissipates energy, so that the relative velocity decreases. The conventional distinction between HRCs and GMCs is that in the latter case, the remnants are gravitationally bound.

\begin{figure*}
	\centering
	\includegraphics{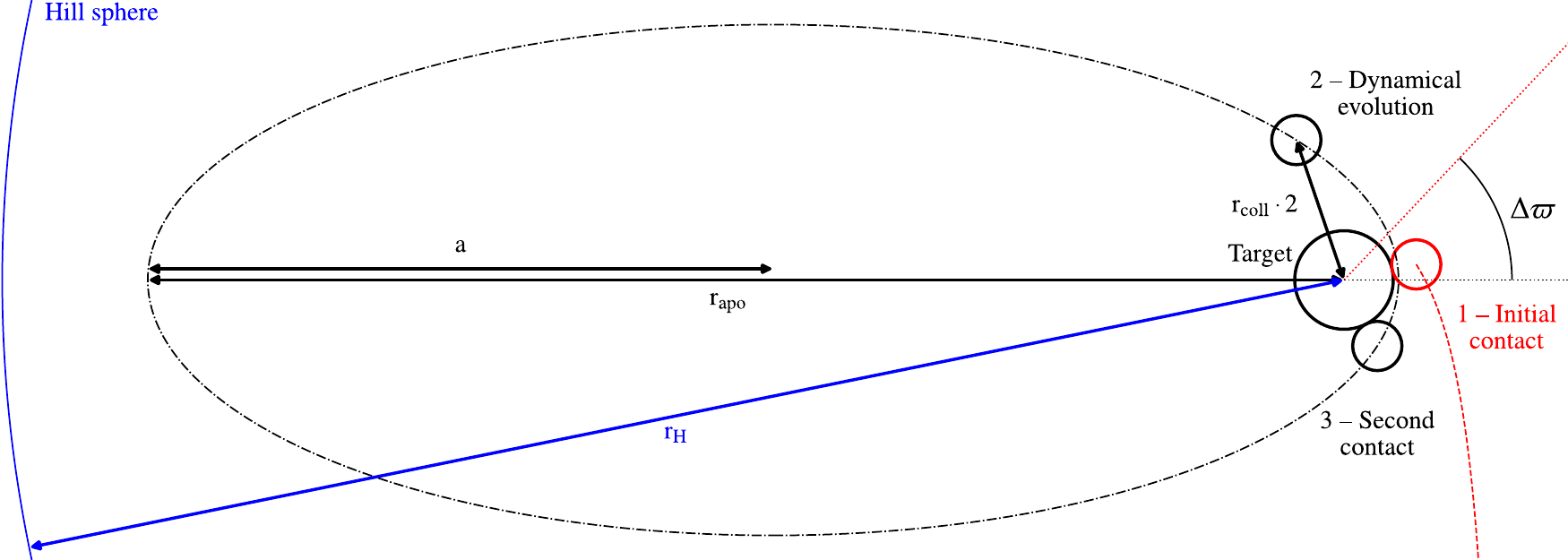}
	\caption{Dynamics of a graze and merge collision, as seen from the target's reference frame. The projectile, coming in on a hyperbolic orbit from the bottom, shown with a dashed red line, whose reference axis is shown with the dotted red line collides with the target at a shallow impact angle (point 1). After the initial collision, a second body (composed mostly from the projectile) remains on an elliptical orbit (point 2), and collides again after one orbit (point 3), resulting in the final merger. The angle $\Delta\varpi$ denotes the shift of the argument of pericenter between the initial hyperbolic orbit and the subsequent elliptic orbit. The blue arc shows an arbitrary Hill radius in case the collision happens when the targets orbits a third massive body. In the present situation, we have $\rapoH=r_\mathrm{apo}/r_\mathrm{H}\simeq0.9$.}
	\label{fig:diag-gnm}
\end{figure*}

The dynamics of such an encounter is sketched on Figure~\ref{fig:diag-gnm}. In addition to energy dissipation, the encounter leads to a shift of the orientation of the orbits, i.e. the argument of pericenter, as the collision happens before the bodies reach their pericenter. This shift depends mainly on the impact angle, as more head-on collision will lead to deeper penetration while in barely grazing events the bodies have a reduced contact surface, with a corresponding less important interaction and energy dissipation.

\begin{figure*}
	\centering
	\includegraphics{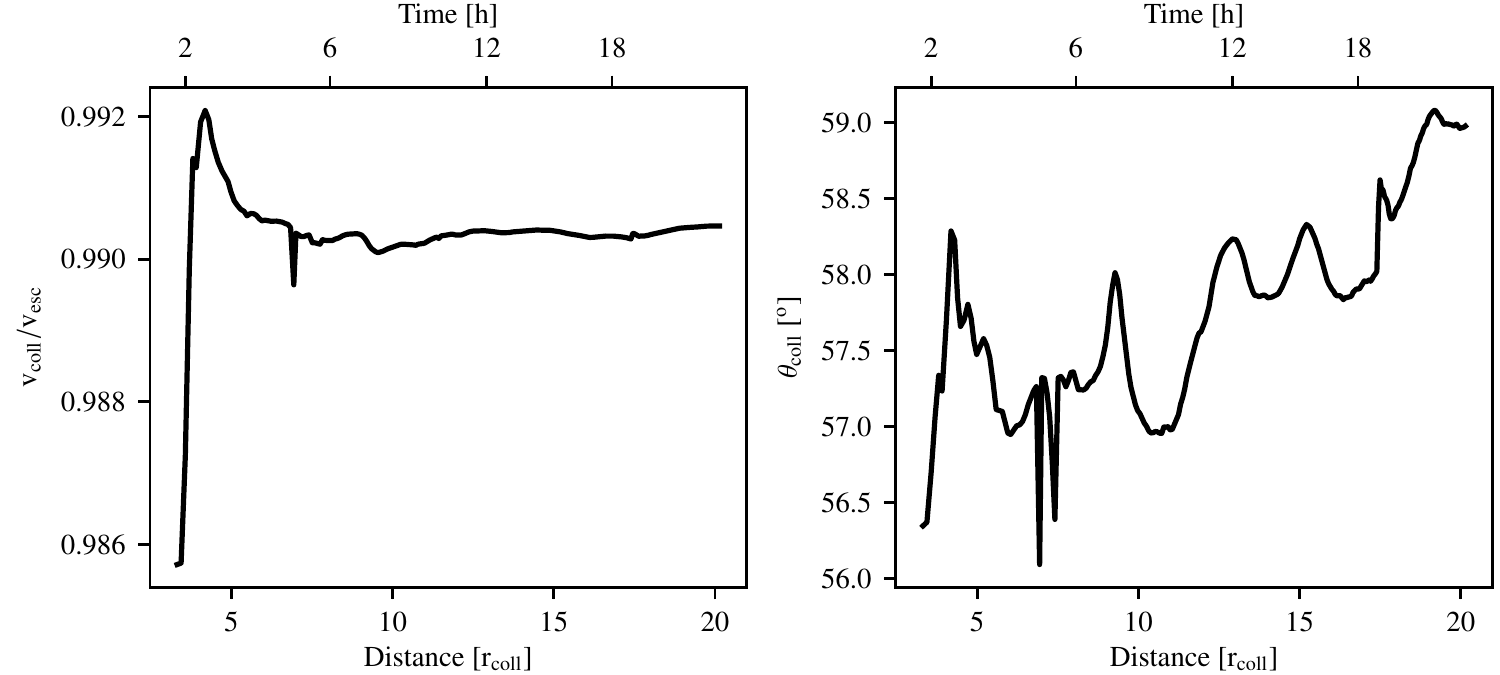}
	\caption{Properties of relative motion after an initial encounter where the two bodies remain bound; i.e. a GMC. The left panel shows the impact velocity, calculated assuming energy conservation, and the right panel the impact angle, calculated assuming angular momentum conservation. The corresponding time since initial encounter is show at the top.}
	\label{fig:gmc-params}
\end{figure*}

Processes involved during collision, and the deformation of the bodies changing their gravitational potential, can result in a trajectory that is not well represented by a simple two-body problem. So, to check our assumption of separability of the different stages of a GMC, we examine one Smoothed Particle Hydrodynamics (SPH) collision simulation part of the \citetalias{2019ApJEmsenhuberA} study, with a target mass $m_\mathrm{tar}=\SI{0.9}{M_\oplus}$, a projectile mass $m_\mathrm{proj}=\SI{0.2}{M_\oplus}$, a impact velocity $v_\mathrm{coll}/v_\mathrm{esc}=1.1$ and an impact angle $\theta_\mathrm{coll}=\ang{52.5}$. This collision was performed in the target-projectile frame, without external perturbation. For this event, we show in Figure~\ref{fig:gmc-params} the orbital properties as function of separation of the two main remnants, given in terms of the collision distance, that is the sum of their radii. To determine the properties of the remnants, we use the scheme as presented in \citetalias{2019ApJEmsenhuberA}, that is a friend-of-friend search, with an improvement described in Appendix~\ref{sec:mix} to take into account the low-density material that is otherwise disregarded.

We find that once the bodies are separated by about six times their mutual radii, the relative motion is well represented by a two-body problem; the two quantities shown on Figure~\ref{fig:gmc-params} being constants of the problem. The impact velocity barely changes after that time while the expected impact angle varies within two degrees. If perturbations due to other bodies are not significant below this distance, then a GMC can be separated into different stages without incurring a loss of accuracy.

\begin{figure}
	\centering
	\includegraphics{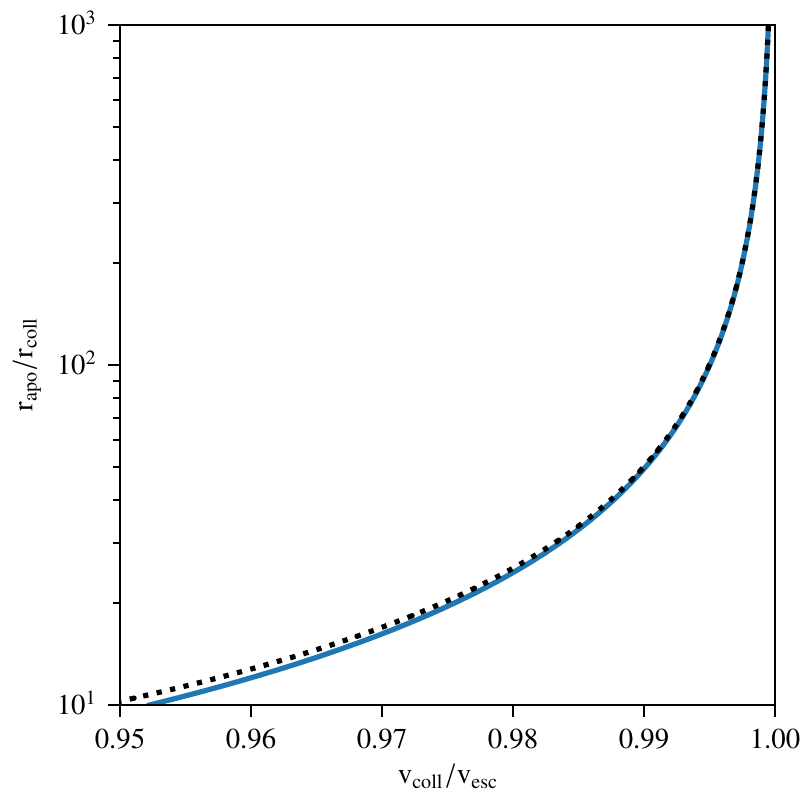}
	\caption{Impact velocity required to achieve a certain apocenter, given in terms of the collision radius. The blue line is computed following the same method described in Section~\ref{sec:methods} while the dotted black line denotes the maximum reachable distance assuming energy conservation (equivalent to a perfect head-on case).}
	\label{fig:dist-vel}
\end{figure}

The end state, and particularly the end velocity determines the range reached by the bodies during the post-encounter orbit. We show the impact velocity depending on the reached distance in Figure~\ref{fig:dist-vel}, for a \ang{60} collision similar to the one shown before (but actually computed using the procedure described in Section~\ref{sec:methods}), and a theoretical calculation based solely on energy conservation (which would be valid for head-on orbits), and given by
\begin{equation}
\frac{v_\mathrm{coll}}{v_\mathrm{esc}}=\sqrt{1-\frac{r_\mathrm{coll}}{r_\mathrm{apo}}},
\label{eq:vel-dist}
\end{equation}
with $v_\mathrm{coll}$ the velocity at initial contact, $v_\mathrm{esc}$ the mutual escape velocity, and $r_\mathrm{coll}=r_1+r_2$ the separation at initial contact.

The two curves match pretty well, indicating that there is only a weak dependence of the orbit geometry on the apocenter. It can also be noted that the range of velocities required to obtain wide orbit is narrow; as an orbit reaching an apocenter of 10 times the sum of the body radii, the result is an impact velocity of $v_\mathrm{coll}/v_\mathrm{esc}\simeq0.95$, while if the same ratio is 100, it results in $v_\mathrm{coll}/v_\mathrm{esc}\simeq0.99$. The situation with $v_\mathrm{coll}/v_\mathrm{esc}=1$ is the limiting case of a parabolic orbit.

\section{Collisions around a central body}
\label{sec:cb}

We continue the process of a GMC by looking at the orbit after the initial encounter. Here we assume a case that results in two remnants of mass $m_1$ and $m_2$ (with $m_1>m_2$) occurs around a central body of mass $m_0$.
To first order, the zone of influence of the largest remnant with respect to the central body is the Hill sphere, whose radius is given by
\begin{equation}
r_\mathrm{H}=a_1\sqrt[3]{\frac{m_1}{3m_0}},
\end{equation}
where $a_1$ is the semi-major axis of the orbit of $m_1$ about $m_0$. If we assume that the largest remnant is on an almost circular orbit around the central body, then Hill radius remains almost constant in time. In case of weakly bound GMC, the orbit of the second remnant is highly elliptical, with an apocenter that we can express in terms of the Hill radius with $\rapoH=r_\mathrm{apo}/r_\mathrm{H}$. If we further assume that the apocenter of the second remnant is much larger than the bodies radii, then $a_2\approx r_\mathrm{apo}/2$ (shown as $a$ and $r_\mathrm{apo}$ on Figure~\ref{fig:diag-gnm}), and we can compute some properties. For instance, the ratio between the orbital periods is
\begin{equation}
\frac{T_2}{T_1}=\frac{2\pi\sqrt{a_2^3/\mu_1}}{2\pi\sqrt{a_1^3/\mu_0}}\approx\sqrt{\rapoH^3\frac{\mu_0}{8\mu_1}\frac{m_1}{3m_0}}\approx\sqrt{\frac{\rapoH^3}{24}}
\label{eq:period-ratio}
\end{equation}
with $\mu_\imath=G(m_\imath+m_{\imath+1})\approx Gm_\imath$ for $\imath=0,1$ the standard gravitational parameters, where the assumption that the mass of the primary dominates. This assumption is in practice incorrect for $\mu_1$, but nevertheless permits to have a good idea of the scale. We obtain here that the period ratio of the unperturbed orbits is scale-independent. Therefore, the relative motion of the central body with respect to the orbital plane during the revolution of the two remnants depends only on the orientation and scaled apocenter distance $\rapoH$. We thus expect that the results obtained for one set of bodies to be generalisable to any scale, as long as the assumptions are retained.

The above formula can be easily corrected for when the mass of the second remnant is non-negligible. In this case, $\mu_1$ is to be replaced by its exact value and a factor $1/(\gamma+1)$, with $\gamma=m_2/m_1$, is present inside the square root. In this case, we obtain again a scale-independence, provided that the mass ratio between the two remnants of the first encounter is fixed.

In case the bodies radii are no longer negligible compared to the Hill sphere, the eccentricities will become lower for a given ratio $\rapoH$, as the pericenter is raised; the latter being in the order of $r_\mathrm{coll}$. Another correction to the Equation~(\ref{eq:period-ratio}) is possible; if we define $\rperiH=r_\mathrm{peri}/r_\mathrm{H}\sim r_\mathrm{coll}/r_\mathrm{H}$, then the term $\rapoH$ is replaced by $(\rperiH+\rapoH)$.

\subsection{Prevalence}
\label{sec:prev}

\begin{table}
    \centering
    \caption{Relationship between the physical and Hill radii of various solar system objects}
    \label{tab:hill-phys-rad}
    \begin{tabular}{ccc}
        \hline
        Body & Primary & $r_\mathrm{H}/r_1$ \\
        \hline
        Ganymede & Jupiter & \num{1.2e1} \\
        Titan & Saturn & \num{2.0e1} \\
        Merucry & Sun & \num{9.0e1} \\
        Venus & Sun & \num{1.7e2} \\
        Earth & Sun & \num{2.3e2} \\
        Mars & Sun & \num{3.2e2} \\
        Jupiter & Sun & \num{7.6e2} \\
        Saturn & Sun & \num{1.1e3} \\
        Uranus & Sun & \num{2.8e3} \\
        Neptune & Sun & \num{4.7e3} \\
        Pluto & Sun & \num{6.4e3} \\
        \hline
    \end{tabular}
\end{table}

The outcome of the first encounter gives the apocenter in terms of the collision radius while in the earlier part of this section, we used the apocenter given in terms of the Hill radius. To convert between the two, we need to obtain the relationship between the Hill and physical radii. This ratio is given by
\begin{equation}
\frac{r_\mathrm{H}}{r_1} = a_1 \sqrt[3]{\frac{4\pi}{9}\frac{\rho_1}{m_0}},
\label{eq:rad-ratio}
\end{equation}
with $r_1$ the physical radius of the body $m_1$ and $\rho_1$ its bulk density. It is independent of the mass $m_1$. The distance of the bodies to the primary, $a_1$, is the most important parameter. Remnants from collisions in the inner part of the system will require a lower relative velocity to reach a sizable fraction of the Hill sphere. On the other hand, only a narrow range of relative velocities is able to reach a fraction of the Hill radius.

We provide a few examples of this value for Solar System bodies in Table~\ref{tab:hill-phys-rad}. We note that the Jupiter or Saturn systems are where bodies whose radius are quite comparable with their Hill radius. The planets themselves have a physical radius much smaller than their Hill sphere. The most distant objects, such as Pluto, have a Hill radius many thousand times greater than their physical radius.

To compare these values with the discussion of the previous section, we further need to relate the collision radius, $r_\mathrm{coll}=r_1+r_2$, to $r_1$. For instance, if we assume both bodies have roughly the same bulk density and take a similar situation than the hydrodynamical simulation presented in the last section with $\gamma=m_2/m_1\approx0.2$, then we can estimate $r_2/r_1\approx0.6$, hence $r_\mathrm{coll}\approx 1.6 r_1$. With this is mind, we can get that in the case of Ganymede, a collision resulting in a relative velocity in the range $0.93 \lesssim v_\mathrm{coll}/v_\mathrm{esc}<1$ will have bodies that are initially bound reaching a distance larger than the Hill radius, while in the case of Pluto this is restricted to $1-\num{1.3e-4}\lesssim v_\mathrm{coll}/v_\mathrm{esc}<1$.

\section{Methods}
\label{sec:methods}

As we are interested in the dynamical effects occurring the intermediate stage of a GMC (i.e. the orbit in between the encounters), we will focus on the dynamical part. Hence we do not perform hydrodynamical simulations of collisions; rather our initial conditions represent the likely outcome of the first encounter of a GMC. We begin with a set of bodies that have an orbital configuration as in the point 2 (``dynamical evolution'') of Figure~\ref{fig:diag-gnm}; that is they are initially separated by twice their mutual radii and moving away from each other. The shape of the orbit is determined as follows: the apocenter is set to given fraction of the Hill radius of the latter, and then the pericenter is iterated until the orbit results in pre-determined impact angle of \ang{60} if no external perturbation were present, a value similar to the end result of the simulation discussed in Section~\ref{sec:gmc} (see right panel of Figure~\ref{fig:gmc-params}).

The dynamical evolution uses the \textit{Mercury} \textit{N}-body code \citep{1999MNRASChambers}. Mercury uses a symplectic integrator scheme \citep[see e.g.,][for a review]{1992ActaNumSanzSerna}. The \textit{N}-body problem is not solvable analytically for more than two bodies. It is nevertheless possible to solve the equations of motion for a problem similar to the original one. The basic principle is to split the Hamiltonian of the system,
\begin{equation}
    \mathcal{H}=\sum_{i=0}^{N-1}\frac{p_i^2}{2m_i}-G\sum_{i=0}^{N-1}m_i\sum_{j=i+1}^{N-1}\frac{m_j}{\Delta x_{ij}},
\end{equation}
into distinct pieces that are each solvable analytically. Here, $\mathbf{p}_i=m_i\mathbf{v}_i$ is the momentum, $\Delta x_{ij}=|\mathbf{x}_i-\mathbf{x}_j|$ the distance between the bodies $i$ and $j$, $G$ the gravitational constant and $N$ the number of bodies in the system. In the general situation, when the bodies are far apart, the dominant force is the contribution by the central body, and the interactions can be treated as perturbations.

However, when two bodies undergo a close encounter, that approximation no longer holds. The force generated by the other close-by body becomes significant, which results in a loss of precision. To circumvent this problem, \textit{Mercury} uses an hybrid approach to resolve close encounters. Pairs that undergo a close encounter are evolved using a Bulirsch-Stoer algorithm \citep{1980BookStoerBulirsch} to resolve the contributions. The transition between the two methods occurs when the two bodies are closer than 3 Hill radii and it is smoothed as a function of the separation so that the Bulirsch-Stoer method has a higher weight when the bodies are closest. This algorithm is discussed in details in \citet{1999MNRASChambers}.

\begin{figure}
	\centering
	\includegraphics{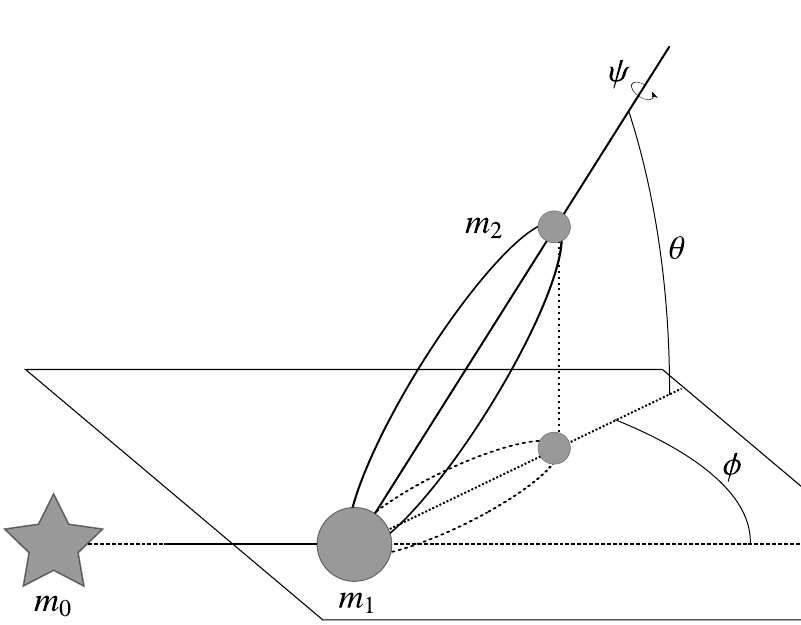}
	\caption{Definition of the different angles. The plane denotes denotes the orbital planet of $m_1$ about $m_0$; $\phi$ is measured in that plane.}
	\label{fig:diag-orientation}
\end{figure}

To obtain distributions of possible outcomes, we adopt a Monte Carlo procedure reminiscent of \citetalias{2019ApJEmsenhuberA}. For every couple of orbital configuration and distance to the star, we perform a series of \num{10000} dynamical evolution runs. The largest remnant is assumed to be on a circular orbit around the central star, while each run has a different orientation of the orbit of the second remnant, $m_2$. The variables $\phi$ and $\theta$ describe the direction in space of the apocenter, and are chosen so that the values are uniformly distributed on a sphere. A third variable, $\psi$, denotes the rotation about the major axis of the ellipse and has an underlying uniform distribution. The overall scheme provides the equivalent of a random distribution, which is consistent with the results obtained in \citetalias{2019ApJEmsenhuberA}.

The definition of $\phi$ and $\theta$ has been changed from \citetalias{2019ApJEmsenhuberA}: $\phi$ represents the direction of the apocenter of the smaller body, in the orbital plane of $m_1$ around $m_0$, with $\phi=0$ denoting an orbit away from the central body, increasing first in the direction of orbital motion. $\theta=0$ has been changed to represent an apocenter in the same orbital plane. A graphical representation of these angles in provided in Figure~\ref{fig:diag-orientation}.

\section{Results}
\label{sec:res}

We begin by applying the method to a case where the Hill sphere is much larger than the bodies. We select $m_0=\SI{1}{\msun}$, $m_1=\SI{1}{\mearth}$, $m_2=\SI{0.1}{\mearth}$ and $a_1=\SI{1}{AU}$. These conditions are similar to the hydrodynamical simulation and dynamical evolution models performed in \citetalias{2019ApJEmsenhuberA}. In the case of the Earth and Sun, the ratio between the Hill radius and the semi-major axis is $r_\mathrm{H}/a\simeq\num{1e-2}$, so that the ratio with Earth's radius is $r_\mathrm{H}/r_\oplus\simeq\num{230}$. This high ratio makes that any orbit that will be modeled is highly eccentric, with $e>0.95$. Hence, a slight change in the orbital energy shifts considerably the apocenter. Thus, it is extremely difficulty to obtain hydrodynamical models that come up with the desired relative velocity, as small changes in the initial impact velocity or angle will lead to quite different apocenter, or unbound bodies.

\subsection{Time between impacts}

\begin{figure}
	\centering
	\includegraphics{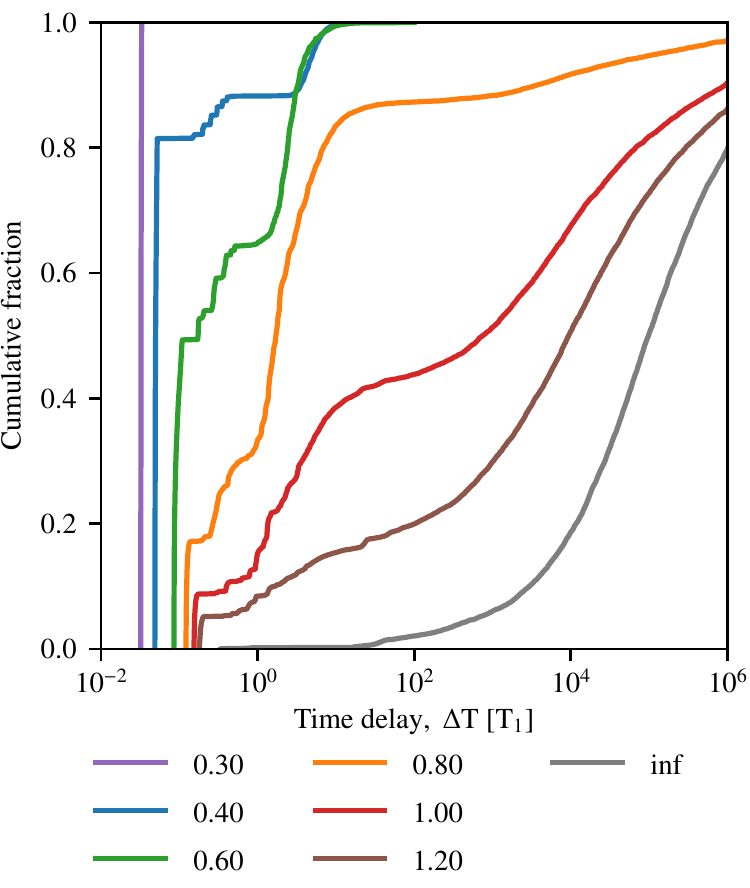}
	\caption{Cumulative distribution of the delay between successive collisions ($\Delta T$) for different values of the scaled apocenter distance $\rapoH$, as provided in the legend for the case of an Earth-like target at \SI{1}{\au} from a Sun-mass star. The last case (in grey) is the limiting case with $e=1$. The time is given in terms of the orbital period of the target around the Star, $T_1=\SI{1}{yr}$.}
	\label{fig:time-series}
\end{figure}

First, we analyse the delays between the successive encounters. The cumulative distributions for various series of dynamical evolution are shown in Figure~\ref{fig:time-series}. The onset of each curve denotes the time required to perform one orbit, which is in good agreement with equation~(\ref{eq:period-ratio}). For the shortest-period set shown here, $\rapoH=r_\mathrm{apo}/r_\mathrm{H}=0.3$, we note that all pairs collide again after a single orbit, whereas when looking at the next series with $\rapoH=0.4$, we observe that about 20\% of the cases do not return after the first orbit. For those, we note that some return within a year (i.e. one orbital period of the target around the star), then a plateau occurs and the remaining collide again within 100 years. Up to $\rapoH\simeq0.6$ we obtain the same general picture, with the fraction of further collisions occurring after a single orbit lowering to less than 50\%. For greater apocenters, we to obtain case that take longer than $\num{1e2}T_1$ to return, and even past $\num{1e6}T_1$, which is when dynamical evolution ends.
When the apocenter would be located at the same distance than the Hill radius, then only about 10\% of realisations lead to a second collision after a single orbit, while most take longer than $\num{1e2}T_1$ to collide again. We nevertheless obtain that more than 90\% of the cases collide within $\num{1e6}T_1$. With the limiting case of a parabolic orbit, we have just a few percent of collision occurring before $\num{1e2}T_1$, but still a return rate of more than 80\% within $\num{1e6}T_1$. The median time is on the order of $\num{1e5}T_1$, which is similar to the results from \citetalias{2019ApJEmsenhuberA}. In this sense, we have the same limiting case either from GMCs that are weakly bound, or HRC that are weakly unbound, hence the transition from the GMC to HRC regime is smooth.

\begin{figure}
	\centering
	\includegraphics{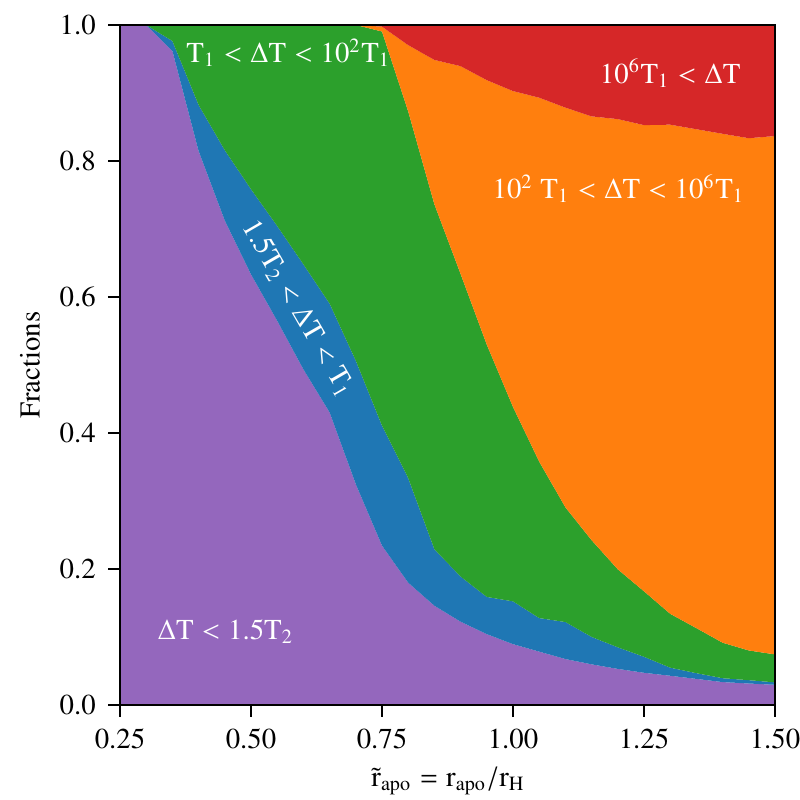}
	\caption{Fraction of time delays as function of the scaled apocenter distance $\rapoH=r_\mathrm{apo}/r_\mathrm{H}$ for the case of an Earth-like target at \SI{1}{AU} from a Sun-mass star, without any other bodies present.}
	\label{fig:times}
\end{figure}

To provide a better overview of the delay between collisions, we show in Figure~\ref{fig:times} a breakdown of the time delay between impacts versus the scaled apocenter distance $\rapoH$ in the two-body problem. The time delay $\Delta T$ is divided into multiple categories: lower than $\num{1.5}T_2$ (purple) are the ones that collide soon after a single orbit; up to $1T_1$ (blue) are the ones that return after multiple orbits but before the target performs a complete revolution around the central body; and then those that return after one, one hundred, or a million times $T_1$, categories chosen so that their boundaries lie near plateaus observed in the cumulative distributions shown in Figure~\ref{fig:time-series}.

The same features that were discussed for the cumulative distributions also appear on this plot. In addition, we can note that the fraction of events that return after a single orbit, but before a revolution of the target about the central star is never preponderant.

The change to regime where the return collision happens after a single orbit, and after $100T_1$, occur over short variation of the apocenter. In the former case, virtually all returns occur on the first orbit for $\rapoH<0.35$, while this fraction is down to 62\% for $\rapoH=0.50$. There are virtually no pairs surviving more than $100T_1$ for $\rapoH<0.75$, while this becomes the principal outcome for $\rapoH\approx0.95$.

In summary, the usual picture of a GMC, with the second collision occurring after a single orbit, is valid only when the apocenter of the orbit remains within $0.3r_\mathrm{H}$. Beyond that value, the strength of the interaction with the central star is enough to modify the orbit so that the objects are not crossing on the first pericenter passage. Up to $0.6r_\mathrm{H}$, all pairs collide further within $100T_1$, so that even though the time span increases, the overall properties of GMCs are retained.

\subsection{Impact properties}

In a standard GMC, the properties of the second collision are set by the end state of the first event. However with the presence of other bodies, this assumption might no longer hold. To check for this, we analyse different impact properties.

\subsubsection{Impact velocity}

\begin{figure}
	\centering
	\includegraphics{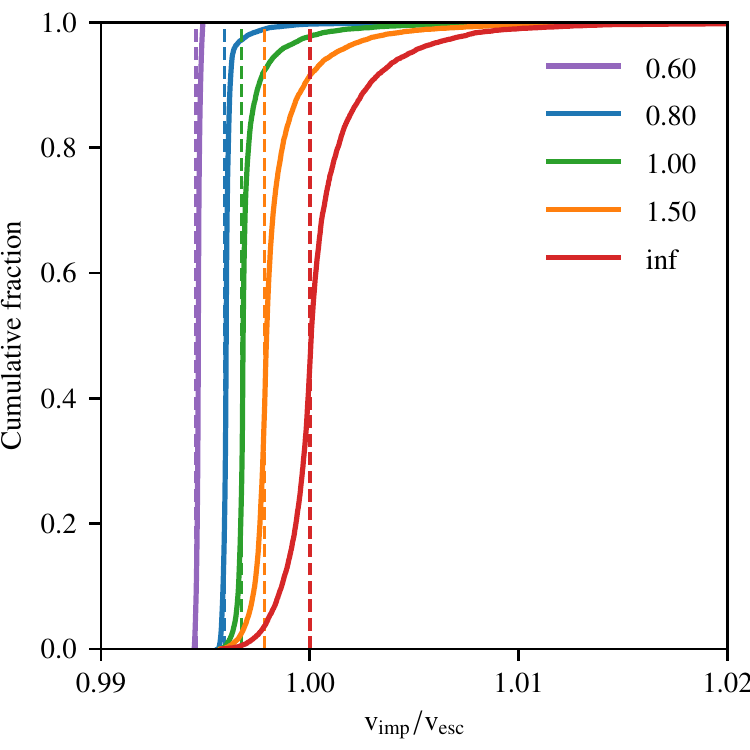}
	\caption{Distribution of impact velocities for the different dynamical evolution series varying as a function of the scaled apocenter distance $\rapoH$ as given in the legend. The expected value for each case if there were no perturbations is given by the vertical dashed lines, whose value is computed with Equation~(\ref{eq:vel-dist}). Note that the colors are for different sets than in Figure~\ref{fig:time-series}.}
	\label{fig:vel-dist}
\end{figure}

We begin with the impact velocity, for which we provide the distributions for different values of the apocenter distance $\rapoH=r_\mathrm{apo}/r_\mathrm{H}$ in Figure~\ref{fig:vel-dist}. Overall, we observe only small variations from the expected values obtained from Equation~(\ref{eq:vel-dist}), and only for relatively distant apoceneters. In the case with $\rapoH=0.60$, we notice almost no difference from the expected value, even though the previous section showed that only about half of the cases collide again after a single orbit. For higher apocenters, some collisions return faster than the mutual escape velocity. However, the highest impact velocity obtained is $v_\mathrm{coll}/v_\mathrm{esc}\simeq1.033$ (for the limiting case where the initial orbit is parabolic); not much higher that the original value $v_\mathrm{coll}/v_\mathrm{esc}=1$.

Thus the impact velocity is barely affected by the presence of the central body. The changes observed here are lower than the expected velocity decrease by a grazing collision, which are on the order of $\SI{0.1}{v_\mathrm{esc}}$ for collisions with an impact angle of \ang{60} \citepalias{2019ApJEmsenhuberA}. Thus the velocity changes in GMCs will be dominated by the encounters rather than the evolution that occurs in between.

\subsubsection{Impact angle}

\begin{figure}
	\centering
	\includegraphics{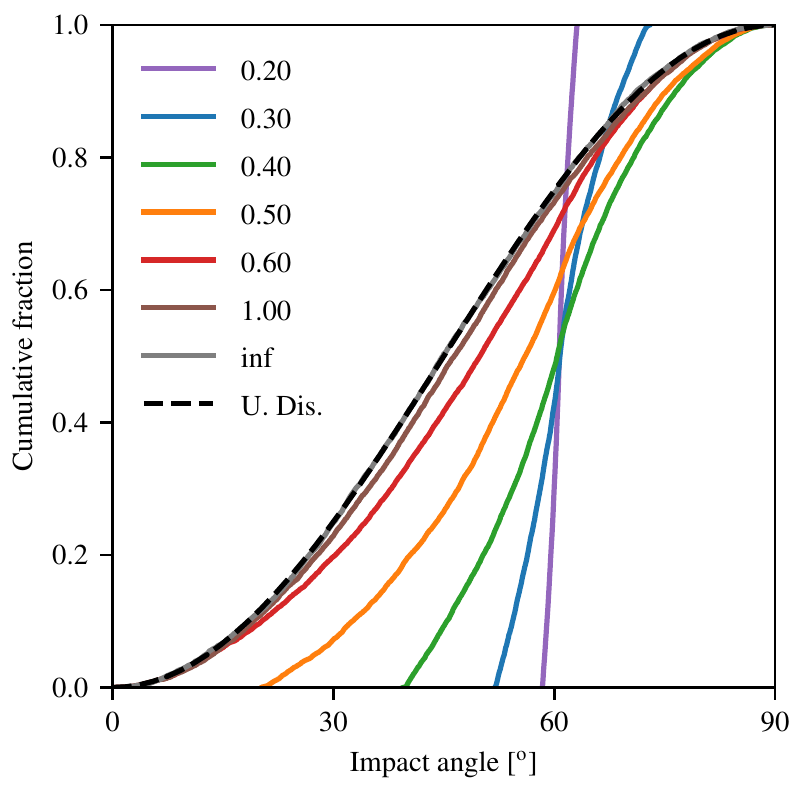}
	\caption{Distribution of impact angle for the different dynamical evolution series varying the as function of the scaled apocenter distance $\rapoH$, as given in the legend. The initial conditions are such that without perturbation, the return impact angle would be \ang{60}. Note that the colors are for different sets than in Figures~\ref{fig:time-series} or~\ref{fig:time-series}. The dashed black line represent the expected distributions for orbits uniformly distributed in space \citep{1962BookShoemaker}.}
	\label{fig:angle-dist}
\end{figure}

The distributions of impact angles for different sets of dynamical evolution with various scaled apocenter distances $\rapoH=r_\mathrm{apo}/r_\mathrm{H}$ are provided in Figure~\ref{fig:angle-dist}. In the case where the orbit is not affect by the presence of the central body, then we expect all subsequent collisions to occur at \ang{60}, following our initial conditions. This is almost the case for the set with $\rapoH=0.2$, as we see deviations by a few degrees at most. Then, as the apocenter increases, the impact angle of the second encounter oscillates around the expected value with a greater intensity, until some get higher than \ang{90}, meaning that the body misses the target after the first revolution. We can correlate these results with the time between the collisions: in the set with $\rapoH=0.3$, no impact at roughly \ang{90} is observed, hence all pairs collide again after a single orbit, while in the set with $\rapoH=0.4$, we see that the distribution reaches \ang{90} and we observe ``misses'', with returns occurring after multiple orbits.

With more distant apocenters, we note that the the impact angle distribution tend toward the expected distribution if orbits were uniformly distributed in space, following \citet{1962BookShoemaker}. This shift indicates that the initial conditions have a lesser importance on the properties of the subsequent collision. For instance, in the series with $\rapoH=1.0$, the distribution is similar to the uniform one, though not compatible: the \textit{p-value} from a Kolmogorov-Smirnof (KS) test is on the order of \num{e-13}. The story with the limiting case with a parabolic orbit is different. Though the overall distribution looks similar to the previous one, the \textit{p-value} is 0.97, meaning that there is no reason to assume it comes from another underlying distribution than the uniform one.

The shift towards the uniform distribution has implications for scenarios built on the general properties of GMCs. For instance, in the \citet{2013IcarusAsphaug} scenario for the formation of Saturn's middle-sized moons, the giant impact ultimately forming Titan provides angular momentum in excess that what can be sustained by a single body \citep{1969BookChandrasekhar}. Hence, a part of the angular momentum is carried away by the material that will form the other moons. As these collisions are grazing (as are GMCs in general), if the impact angle of the subsequent relates more to a uniform distribution than from the end stage of the initial collision, then the average impact angle decreases. This reduces in turn the angular momentum of the merger collision, potentially reducing the amount of material that needs to be ejected in the process.

\subsubsection{Alignment}

\begin{figure}
	\centering
	\includegraphics{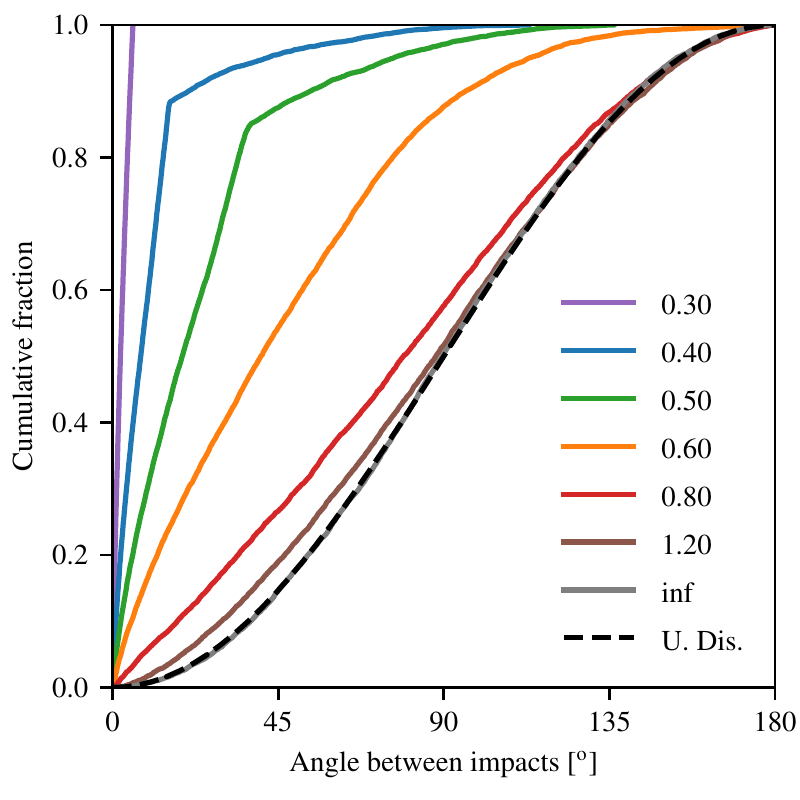}
	\caption{Distribution of angle between symmetry planes of the successive collisions for the different dynamical evolution series varying the as function of the scaled apocenter distance $\rapoH$, as given in the legend. Note that the colors are for different sets than in Figure~\ref{fig:time-series} and Figure~\ref{fig:angle-dist}. The dashed black line represent the expected distributions for orbits uniformly distributed in space.}
	\label{fig:orientation-dist}
\end{figure}

\begin{figure}
	\centering
	\includegraphics{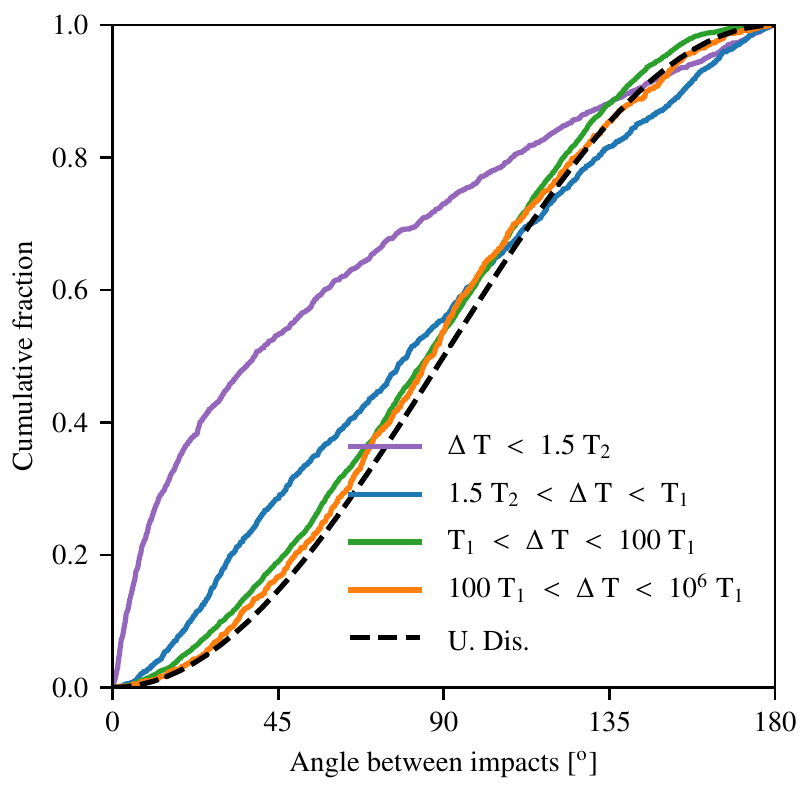}
	\caption{Distribution of angle between symmetry planes of the successive collisions for the dynamical evolution series with $r_\mathrm{apo}/r_\mathrm{H}=0.80$, and broken down by time delays between the collisions. Colors are identical to the categories provided in Figure~\ref{fig:times}. The dashed black line represent the expected distributions for orbits uniformly distributed in space.}
	\label{fig:orientation-time}
\end{figure}

Since we have a succession of two collisions, we can derive a relative orientation. We define the relative orientation (or alignment) by the angle formed by the specific angular momentum vector of the initial orbit and the same at initial contact of the second collision. In the case of unperturbed GMCs the successive events occur in the same plane due to symmetry and angular momentum conservation, while for HRR, we expect no correlation, thus the relative orientation has a distribution compatible with a uniform distribution in space \citepalias{2019ApJEmsenhuberA}.

The distribution of alignments for different apocenter distances is provided in Figure~\ref{fig:orientation-dist}. In case the bodies remain well within the Hill sphere, up to $r_\mathrm{apo}/r_\mathrm{H}=0.3$, then the two impact are well aligned, and the shift between the impact planes remains lower than \ang{10}. However, as soon as some of the pairs do not collide again after a single orbit, we obtain events with a much greater offset, with values up to the order of \ang{90}. Further on, the distribution tends toward the theoretical expectation from uncorrelated events. At $r_\mathrm{apo}/r_\mathrm{H}=0.8$, the mean of the obtained distribution are at \ang{78}, which is not very different from the \ang{90} of the theoretical distribution. So in the case of alignment, GMCs that reach this distance have only a weak memory of the previous event. They relate more to HRR than unperturbed GMCs.

Unlike the impact angle, there is a strong correlation between the offset and the time delay between the collisions. We show this in Figure~\ref{fig:orientation-time} for the case with $r_\mathrm{apo}/r_\mathrm{H}=0.8$. Subsequent collisions that occur after a single orbit are mostly aligned with the prior event, though not as strongly as in cases with lower apocenters. The second category, with delays longer than a single orbit, but within one revolution around the central body are less aligned.

For those two categories, we find that there is an excess of events close to \ang{0} or \ang{180}. For longer delays, the distributions, although not compatible, resemble quite well to the distribution for random orientation. The first one, i.e. for collisions between 1 and 100 revolutions around the central body is still shifted toward more aligned collision by a few degrees, and the later one, for collision after 100 revolutions has a \textit{p-value} of 1.6\%. Even if there is a memory of the first collision, the effect is very weak.

It can be argued that shifting the plane in which the collision occurs requires to change the orientation of the relative angular momentum vector, while changing the impact angle requires an alteration of its magnitude.

\subsection{Correlation between orientation and outcome}

\begin{figure*}
	\centering
	\includegraphics{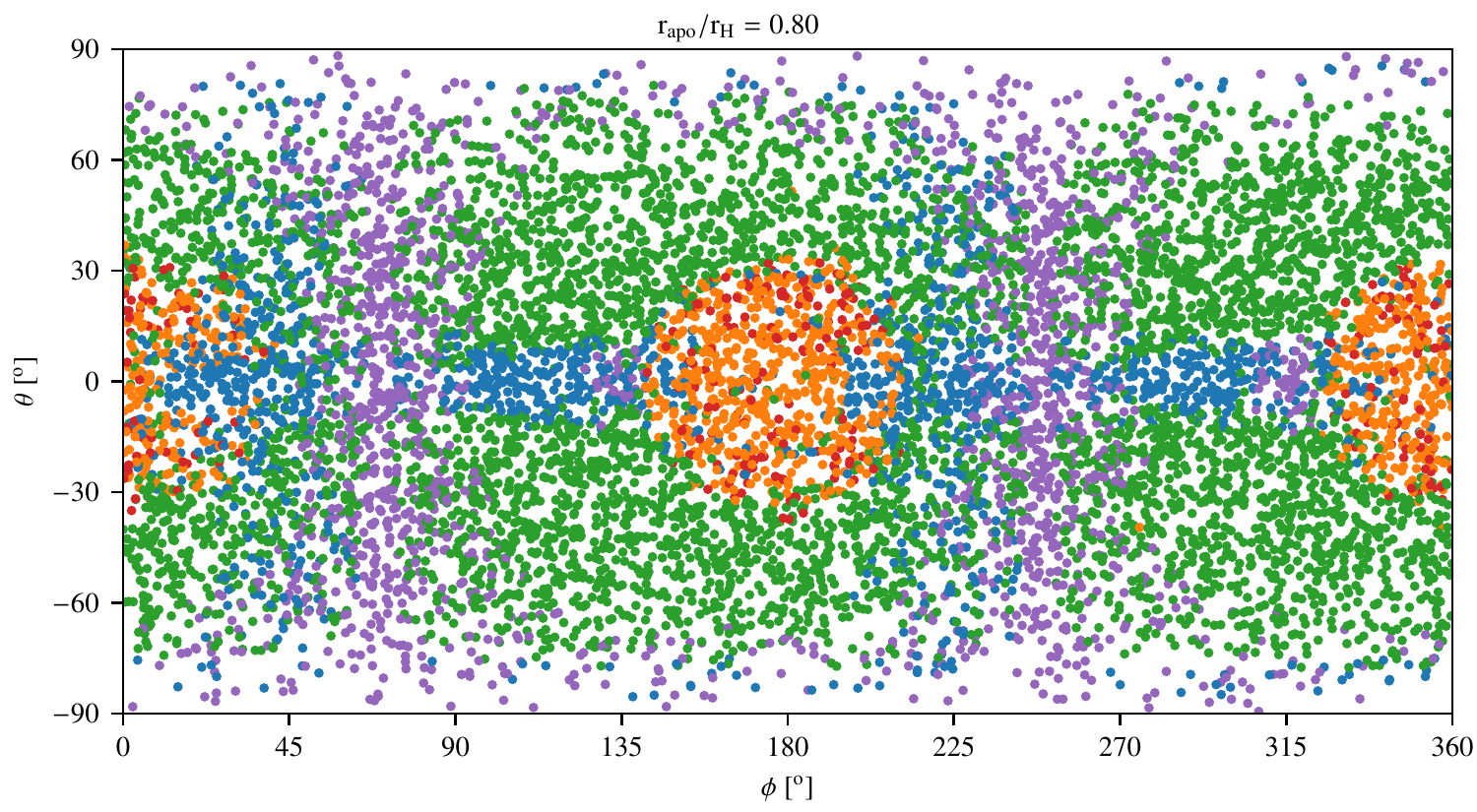}
	\caption{Time delay between successive collisions as function of the scaled apocenter distance $\rapoH=r_\mathrm{apo}/r_\mathrm{H}$ from the two-body problem. Colors are the same as in Figure~\ref{fig:times}: purple for subsequent before $1.5 T_\mathrm{2}$, blue for within one year, green for within 100 years, orange for within 1 Myr and red in other cases. An animated version of this figure is available in the HTML version of this article. The animation spans $\rapoH=0.10$ to $1.50$ in increments of $0.05$. The static representation is provided for $\rapoH=0.80$. For the lowest apocenters, nearly all cases results in a second collision within $1.5\cdot T_\mathrm{2}$ (shown in purple). Starting with $\rapoH=0.75$, extended delays (shown in orange and red) are obtained, first around $(\phi=\ang{0},\theta=\ang{0})$ and $(\phi=\ang{180},\theta=\ang{0})$, until filling the entire space, except for cases near $\phi=\ang{60}$ and $\phi=\ang{240}$. The angles are defined so that the smaller bodies moves towards the central star when $\phi=\ang{180}$ and $\theta=\ang{0}$ (see Figure~\ref{fig:diag-orientation} for the meaning of these angles). Note that the initial conditions have one more angle, $\psi$, which is not shown here.}
	\label{fig:time-conds}
\end{figure*}

We further analyze the relationship between the initial orientation and the type of event. This analysis is similar to the one performed in \citetalias{2019ApJEmsenhuberA}. For reference, in that work we did not find any such correlation, at large or smaller scale. However, the story is different for GMCs that are only slightly perturbed by other bodies, as the number of encounters, and the relative time between collisions lower. Also, we use the post-collision orbit, whereas in the previous work this comparison was performed with the pre-collision orbit.

In the present case, we determine the type of event by the delay between the collisions, which provides a categorization similar to the one shown in Figure~\ref{fig:times}. The results are provided in Figure~\ref{fig:time-conds}. Orbits along the orbital plane of the target around the central body lie in the $\theta\approx\ang{0}$ region, while the upper and lower boundaries of the figure show ``vertical'' orbits, i.e. perpendicular to the orbital plane around the central body. The horizontal axis is defined so that $\phi\approx\ang{0}$ represents orbits where the smaller body goes away the central body, $\phi\approx\ang{90}$ where the same goes towards orbital motion, $\phi\approx\ang{180}$ in the direction of the central body, and $\phi\approx\ang{270}$ against orbital motion; see Figure~\ref{fig:diag-orientation} for a graphical representation.

We find that there is a strong correlation between the delay and the orientation of the orbit. Specific orientations cause both short (less than $\SI{1}{T_1}$; shown in purple and blue on Figure~\ref{fig:time-conds}) and long delays (more than $\SI{e2}{T_1}$; shown in orange and red), while medium delays fill the remainder of the space.

Orbits that have the highest radial excursion towards or against the star ($\theta\approx\ang{0}$ with $\phi\approx\ang{180}$ or $\phi\approx\ang{0}$ respectively) are linked to long delays between collisions. We find that some realisations with similar orientations still return earlier, but these do not follow a clear pattern. For instance, the realisations that exhibit this behavior change for different values of the scaled apocenter distance. On the other hand, orbits that are almost vertical also return after a short time span. The same also applies to apocenters with specific values of $\phi$, but in a direction not exactly towards or against orbital motion. In both cases, we see a shift on the order of \ang{20} to \ang{25} from orbital motion in the positive direction. These situations result in a specific configuration where the overall effect of tidal forces by the central body on the orbit of the two remnants is small, by cancellation of the effects due to the change in distance from the star and non-Keplerian velocity. Hence the resulting orbit, as seen from the rest frame of either body, looking the elongated ellipse that would be expected without perturbation.

We also note that the cases where the subsequent collision occurs past the first orbit, but within a revolution around the central body, lie mostly close to the orbital plane, with a few located near vertical. The first situation has some unique properties, mainly that the bodies never have a large vertical separation. Hence, those are quite close to 2-D systems. So even though the orbits are perturbed by the central body, the collision probability is increased with respect to the general situation, so that the second collision happens on a shorter time scale.

The shape of this plot depends on the scaled apocenter distance. As the value increases beyond $0.3$, the returns taking longer than a single orbit begin to appear around the direction to the central body, and its opposite. The same applies for the cases that escape fully the target, and take more than $\SI{e2}{T_1}$ (orange and red on the figure), and the size of these zone increase, while they remain centered on the direction, and its opposite of the central body. The specific zone of early returns, along the orbital plane and the two values of $\phi$ remains at the same locations, though they become narrower.

The direction shown on the plot is the orientation of the orbit after the initial encounter. Since grazing collisions lead to a shift of the orientation during the encounter (see relevant discussion in Section~\ref{sec:cb} and Figure~\ref{fig:diag-gnm}), this does not immediately relates to the pre-collision orbit. Therefore, there is no immediate way to say whether projectiles coming from a specific direction will return early or late, as the orientation of the impact plane is also random, and will cause a deflection in different directions.

\subsection{Orientation of the subsequent collision}

\begin{figure*}
	\centering
	\includegraphics{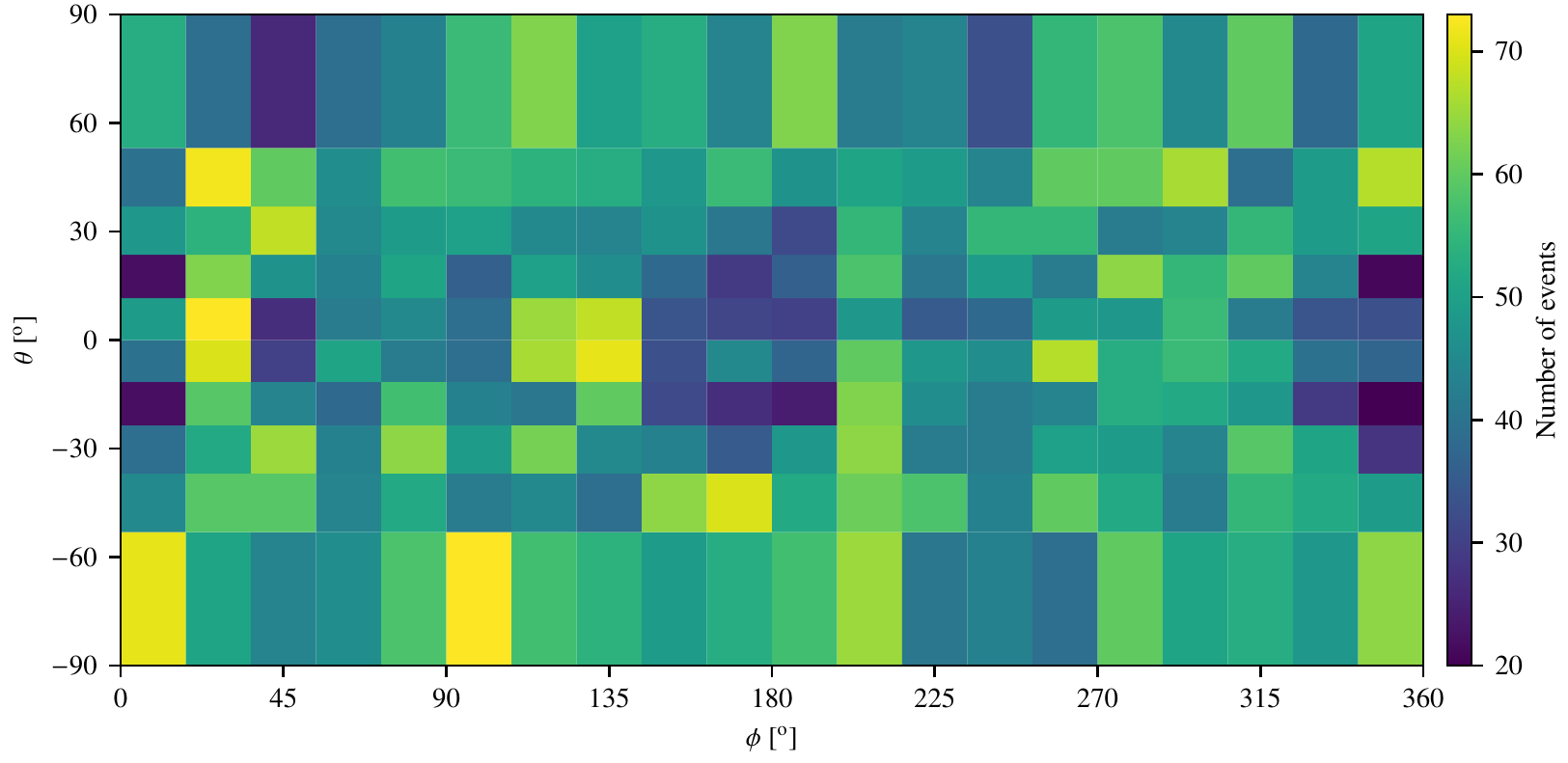}
	\caption{2-D histogram of the orientation of the second collisions for the set with $r_\mathrm{apo}/r_\mathrm{H}=0.8$. The values of $\phi$ and $\theta$ are determined from the orbital motion at the moment the second collision is detected. The position of the bin edges were selected so that each bin has the same probability assuming a uniform distribution in space.}
	\label{fig:second-orientation}
\end{figure*}

The clear spatial segregation of delays has an effect on the possible orientations of the subsequent collision, which can significantly affect the final collisional outcome. Early returns, within a year or so, keep a stronger memory of the orientation of the first collision than the events with greater delays, which tend to behave more like HRC, i.e. by forgetting almost entirely this information \citepalias{2019ApJEmsenhuberA}. This means that there are only few subsequent collisions occurring with orbits aligned toward or against the central body, as the orientation of those will be scattered across the whole space during the dynamical evolution.

To check for this effect, we show the orientation of the subsequent collision in Figure~\ref{fig:second-orientation}, for the series of dynamical evolution performed with $r_\mathrm{apo}/r_\mathrm{H}=0.8$. First we observe a general trend with few events occurring with $\theta\approx\ang{-90}$ or $\theta\approx\ang{90}$; this simply arises from the uniform distribution in space, and is not specific to this situation. On the other hand, we observe that there is an anisotropy in the distribution of $\phi$ for collision occurring near the orbital plane, i.e. with $\theta\approx0$. There are two points with a lower occurrence rate of events near to the direction of the central body, as well as the opposite direction. This confirms the supposition made in the previous paragraph. Nevertheless, it should be noted that these two regions are not free of collisions, and they occur at a greater rate than in the regions close to a vertical alignment. Indeed, we noted in the previous section that even the early returns are not perfectly aligned with the previous collision (see Figure~\ref{fig:orientation-time}).

The location of the highest rate of return collisions is somewhat shifted from the location of early returned observed on Figure~\ref{fig:time-conds}. For the return collisions, we observe maximums around $\phi\approx\ang{30}$, $\ang{130}$, $\ang{210}$, and -- to a lower extend -- around $\ang{280}$. In the map with the initial orientation, we note that returns within an orbital period are located around $\phi\approx\ang{110}$ and $\ang{300}$, as well as similar cases $\phi\approx\ang{45}$ and $\ang{225}$, but only for collisions occurring close to the orbital plane. Furthermore, the locations of the maximums in the return collision orientations close to $\phi\approx\ang{30}$ and $\ang{210}$ overlap with the edge of the region in the initial orientation where the bodies collide again only after an extend period, or not at all. As the latter have only a low chance to return to these orientations, it follows that these maximum are due to early returns from nearby original orientations. We can estimate the change in $\phi$ due to the orbit around the central body while the two collision remnants orbit one another. Using Equation~(\ref{eq:period-ratio}), we obtain that in our situation $T_2/T_1\simeq0.18$, so that a displacement $\Delta\phi\simeq\ang{66}$ is obtained if the alignment of the orbit is not perturbed by the central body.

These observations only apply in a region where the scaled apocenter distance $\rapoH$, is between about 0.5 and a few. For lower values, the orbits are not sufficiently perturbed by the central body, and they all retain the orientation of the initial encounter, while for when the orbit ranges well outside of the Hill sphere, all directions are affected and the orientation follows again a uniform distribution, but without knowledge of the initial encounter. Only in this middle region, some orbits are strongly perturbed while others are not.

The implications of this are that there are still collisions occurring in regions for which return takes a long time (i.e. more than $\Delta T/T_1\gtrsim100$), at least at a greater rate that would be expected from simple considerations. Hence, it is not excluded for collisions to have orientation along the direction of the central body, or the opposite. This is of special significance for collisions that can profit of specific orientation, such as we can expect for, e.g., the proposed scenario proposed in \citet{2013IcarusAsphaug} for the formation Saturn's middle-sized moons.

\subsection{With other bodies}

In \citetalias{2019ApJEmsenhuberA}, we observed that the presence of other bodies affects the return probability, even for when the remnants are barely unbound. In such a situation, we found that the return rate for HRC that results in similar bodies is around 60\% within \SI{20}{\myr}, compared to more than 90\% when only the remnants and the central star are present. This indicates that presence of other solar-system-like bodies should also affect the case where the remnants are initially bound together. We thus extend our study by performing the same dynamical evolution, but with the addition of Mercury, Venus, Mars and Jupiter-like bodies, with their present-day characteristics. The initial location of the planets within their orbit is random, but kept the same across all realisations.

In the cases where the orbit remains well within the Hill sphere, we observe basically no effect due to the presence of the additional bodies. This is the case when $r_\mathrm{apo}/r_\mathrm{H}\leq 0.70$, i.e. when all returns occur with 100 years. For longer periods, the other bodies start to affect the outcomes. At $r_\mathrm{apo}/r_\mathrm{H}=1.00$, the fraction of returns within \SI{1}{\myr} is at 81\% with the additional bodies, compared to 90\% without. The former case was evolved to \SI{20}{\myr}, and if we account for all collisions that occurred during this period, the figure increases to 85\%. During the same period, 7\% of the projectiles collided with other bodies (principally Venus). So this is not only a delay of the return collisions, but also a overall reduction of the probability, much like for HRC \citepalias{2019ApJEmsenhuberA}. In the $r_\mathrm{apo}/r_\mathrm{H}=2.00$ case, we obtain a return probability of 61\% and 70\% within \SI{1}{\myr} and \SI{20}{\myr} respectively. This situation is similar to HRC occurring with no relative velocity at infinity. We then have a transition from something that resembles more to the case without other bodies near $r_\mathrm{apo}/r_\mathrm{H}=1.00$ to something similar to a standard HRC case at $r_\mathrm{apo}/r_\mathrm{H}=2.00$. For reference, we also modeled evolution of situation with more distant apocenters, but with only \num{1000} realizations each time due to the computational burden of this situation (longer delays and higher number of bodies). These other sets are in line with the HRC scenario, meaning that they have lower rate, about 66\% within \SI{20}{\myr} for $r_\mathrm{apo}/r_\mathrm{H}=5$, down to 62\% for $r_\mathrm{apo}/r_\mathrm{H}=100$. Most of the transition thus occurs for $r_\mathrm{apo}/r_\mathrm{H}$ between 0.8 and a few. Past this separation, there is only a weak effect of the distance.

The overall result shows that, at least in this situation, it is the central body that plays a major contribution to perturbation of GMCs. Perturbations start at about $r_\mathrm{apo}/r_\mathrm{H}\sim0.3$ for the central body, and only at greater distances for other bodies, i.e. when second remnant from the collision ranges outside of the Hill sphere of the former, which begins at $r_\mathrm{apo}/r_\mathrm{H}\sim0.8$.
Of course, this might be different with more massive or closer bodies.

\section{Dependency on other properties}

In Section~\ref{sec:cb} we provided an argument that the outcome was primarily depending on the scaled apocenter distance $\rapoH=r_\mathrm{apo}/r_\mathrm{H}$, and we subsequently performed the calculation for a specific configuration, only varying that parameter. Here, we verify this proposition by varying other quantities: the distance between the central body and the largest remnant, $a_1$, the radius of the bodies, and all the object properties, having a Saturn-like system, which is on completely different scale.

Changing the distance between the central body and the largest remnant is actually varying the ratio between the separation at initial contact and the Hill radius; so instead of giving the distance to the star, we define $\rcollH=r_\mathrm{coll}/r_\mathrm{H}$, and refer to the distance using its inverse $\rcollH^{-1}$. For reference, the case analyzed in the previous section had $\rcollH^{-1}\approx153$. To verify that the changes are only due to modification of the shape of the orbit, we also arbitrarily modify the body radii to obtain another method of changing $\rcollH$, while keeping the bodies at the same distance from the central body.

For the Saturn-like system, we set $m_0=1\,M_\text{\saturn}$, $m_1=1\,M_\mathrm{Titan}$, $m_2=0.1\,M_\mathrm{Titan}$, and $a_1=100\,R_\text{\saturn}$. The location of the Titan-like has been artificially increased by a factor five compared to reality. This was chosen to ensure that the ratio between the radius of the body and the Hill radius remains big, as the relationship is given $r_\mathrm{H}/R_\mathrm{Titan}\approx a_\mathrm{Titan}/R_\text{\saturn}$.

Overall, we observe a dependence on the ratio $r_\mathrm{coll}/r_\mathrm{H}$, but not on anything else. For instance varying the distance to the star or the body radii results in the same outcome, as long as the $\rcollH$ parameter remains the same. The same observation can be made when scaling down the whole system to a Saturn-like one, as long as $\rapoH$ and $\rcollH$ are retained. Therefore, the argument we made in Section~\ref{sec:cb} about the scale invariance is mostly confirmed, with a correction for the shape of the orbit. Hereafter, we will only present results obtained with the first method.

For a fixed scaled distance $\rapoH=r_\mathrm{apo}/r_\mathrm{H}$, the most perturbed bodies are for the lowest $\rcollH=r_\mathrm{coll}/r_\mathrm{H}$ values, that is for bodies that much smaller that their Hill sphere. The the eccentricity $e$ is related to ratios $\rapoH$ and $\rperiH$ with $e=(\rapoH/\rperiH-1)/(\rapoH/\rperiH+1)$, where in the case $\rcollH \ll 1$ we can use the approximation $\rapoH/\rperiH \approx \rapoH/\rcollH$, and in the case $\rapoH/\rperiH\gg 1$, we can also use $e\approx1-2(\rperiH/\rapoH)\approx 1-2(\rcollH/\rapoH)$. Note that in the case the orbit is just grazing at the pericenter, i.e. the impact angle is \ang{90}, then $\rcollH=\rperiH$, otherwise $\rperiH<\rcollH$. This means that orbits with a lower value of $\rcollH$ are the most eccentric, and so low eccentricity orbits are more stable than the very eccentric ones. Physically, a small perturbation of a very eccentric orbit will lead to a large shift on the relative position when the two bodies finish their orbit, whereas in the case of a low eccentricity, a larger perturbation of the orbital motion is required to produce a noticeable difference after the completion of the orbit.

\subsection{Effect of the Hill radius}

\begin{figure}
	\centering
	\includegraphics{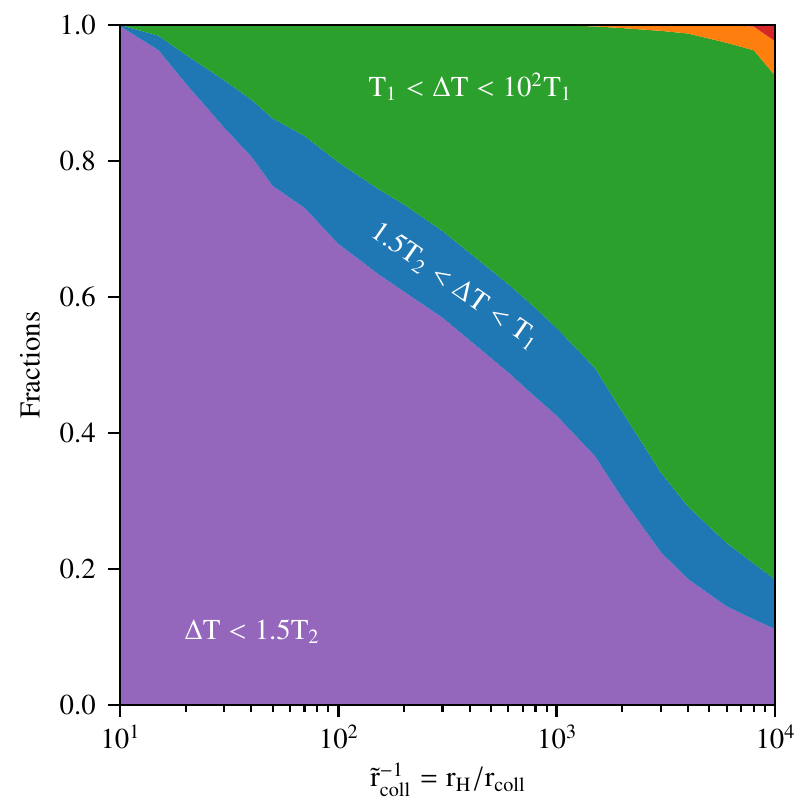}
	\caption{Fraction of time delays ($\Delta T$) as function of the inverse of the scaled physical radius $\rcollH$, for $\rapoH=0.5$. The colors are the same as in Figure~\ref{fig:times}; the two small regions on the upper right are for $10^2T_1<\Delta T<10^6T_1$ (orange) and $10^6T_1<\Delta T$ (red).}
	\label{fig:times-rapo0p50}
\end{figure}

We show in Figure~\ref{fig:times-rapo0p50} the fraction return time delays versus the inverse of $\rcollH$ (i.e. $\rcollH^{-1}=r_\mathrm{H}/r_\mathrm{coll}$) for a fixed scaled apocenter distance $\rapoH=0.50$, broken in the same categories as in Figure~\ref{fig:times}. It can be seen that the general behavior is similar to what is shown in Figure~\ref{fig:times} for values of $\rapoH$ varying between $0.3$ and $0.8$. Here however, the dependence on the parameter is weaker, as the values of $\rcollH^{-1}$ span three orders of magnitude, from \num{e1} to \num{e4}. The other properties, such as the impact angle and the offset between the collision planes, are affected in a similar manner for the same variation of $\rapoH$.

\begin{figure}
	\centering
	\includegraphics{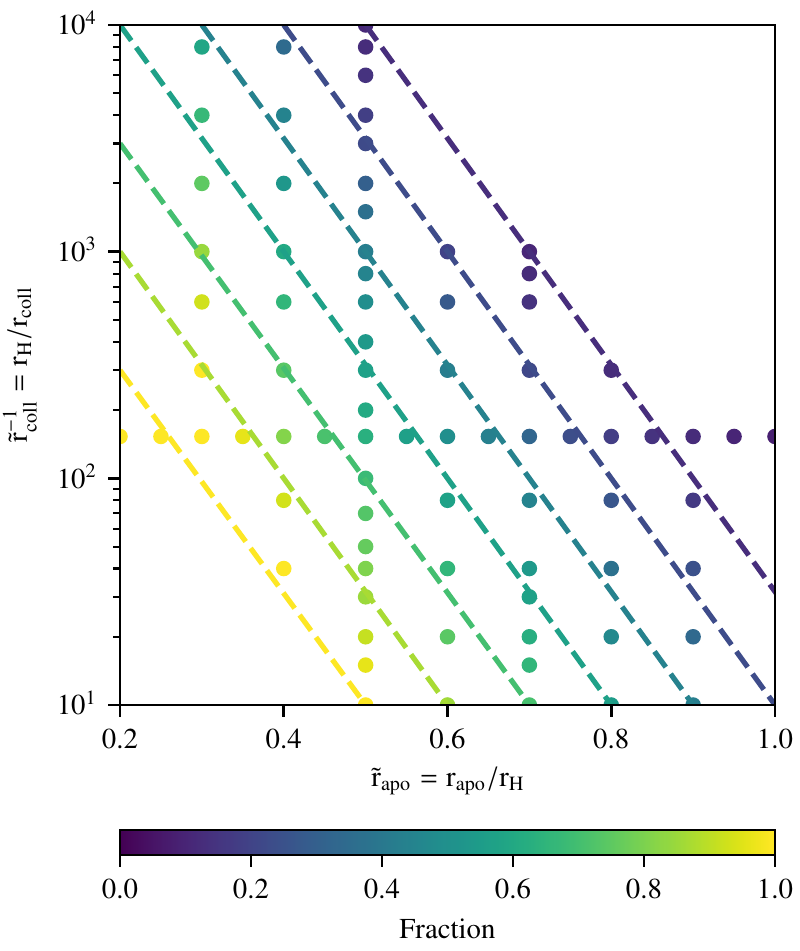}
	\caption{Fraction of runners that return within $1.5T_2$ for each set of dynamical evolution. The dashed lines represent a rough fit (see text); the values for the colors of each of the lines are 1.00, 0.87, 0.70, 0.57, 0.44, 0.23 and 0.13, from the bottom left to the top right.}
	\label{fig:early-map}
\end{figure}

The location of the longest delays in the $\phi-\theta$ space is however different than for greater scaled apocenter distances. When $\rapoH\approx1$, the orientation most likely to have a late return (or none) is when the apocenter is pointed towards or against the central body (see Figure~\ref{fig:time-conds}). When $\rcollH^{-1}$ is large, on the other hand, there is not such a strong correlation. Late returns occur principally for orbits aligned along two directions: with the second body going towards the central body ($\phi\approx\ang{180}$), or behind along the direction of motion ($\phi\approx\ang{270}$). Also unlike the previous situation, these are not symetrical, which means we do not observe the longest delays for second bodies goings away from the central body, or forward in the direction of motion. Further, they are not restraint along the orbital plane, but occur for a wide range of values of $\theta$, expect close to the poles. Another difference is that most of these bodies do not leave the Hill sphere; it is however unclear to us why these alignments tend to produce long delays between impacts.

Since the distribution of time delays is roughly similar whether $\rapoH$ or $\rcollH$ is varied, it is possible to make find a relationship so that the distribution of time delays depends on a single parameter. As an example, when increasing $\rapoH$ from 0.5 to 0.7, we obtain a relatively similar distribution when the factor $\rcollH$ is divided by 10. It is then possible to use a relationship in the form of
\begin{equation}
    k \approx \rapoH-0.2\log_{10}(\rcollH)
    \label{eq:scaling}
\end{equation}
to reproduce the results of Figure~\ref{fig:times} for any value of $\rcollH$, keeping in mind that that figure was generated with $\rcollH^{-1}\simeq153$. To highlight this relationship, we provide in Figure~\ref{fig:early-map} a 2D map of the fraction of returns after a single orbit, i.e. the size of the purple region on Figures~\ref{fig:times} and~\ref{fig:times-rapo0p50}, for various combinations of the scaled apocenter distance $\rapoH$ and body size $\rcollH^{-1}$. The higher resolution data on a horizontal line with $\rcollH^{-1}\simeq153$ and vertical with $\rapoH=0.5$ are the underlying points used for Figures~\ref{fig:times} and~\ref{fig:times-rapo0p50} respectively. The rough relationship between $\rapoH$ and $\rcollH$ is provided with the dashed oblique lines. It can be seen that this fit works relatively well for the region with $\rapoH\gtrsim0.4$. For lower scaled apocenter distance however, the fit underestimates the fraction of early returns.

\subsection{Scaling of the other quantities}
\label{sec:other}

\begin{figure*}
	\centering
	\includegraphics{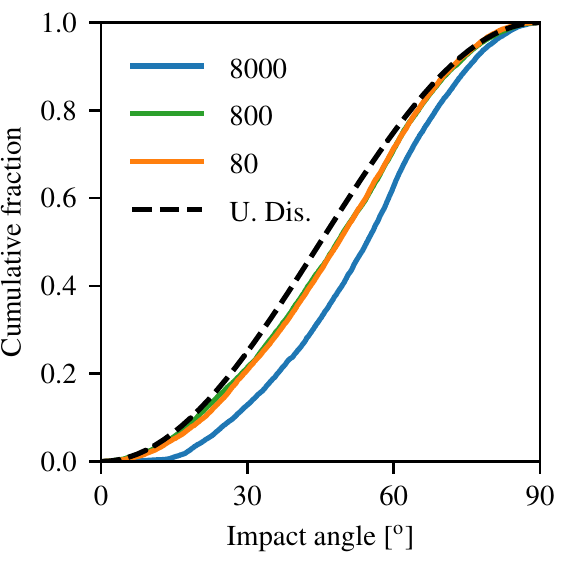}
	\includegraphics{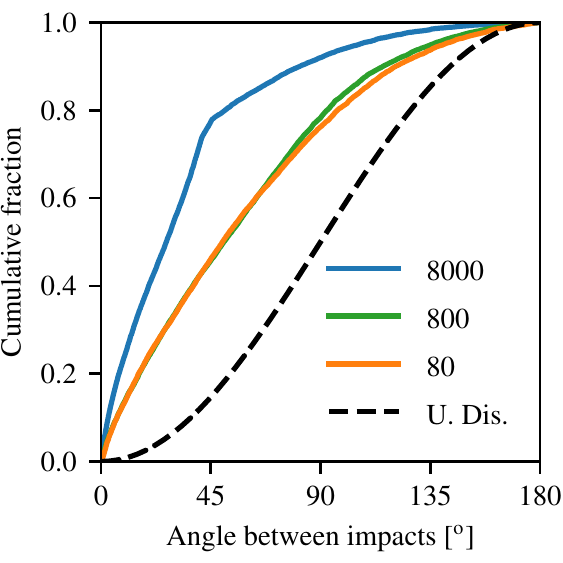}
	\includegraphics{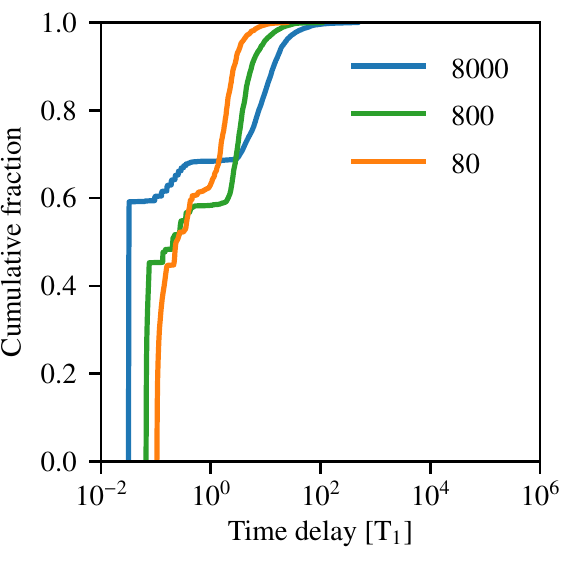}
	\includegraphics{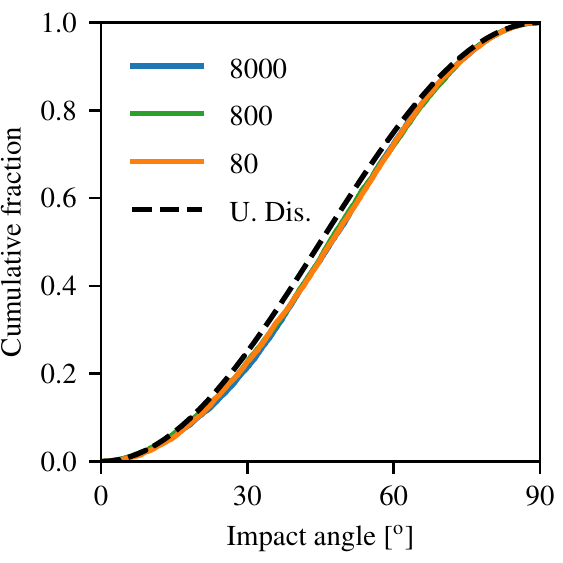}
	\includegraphics{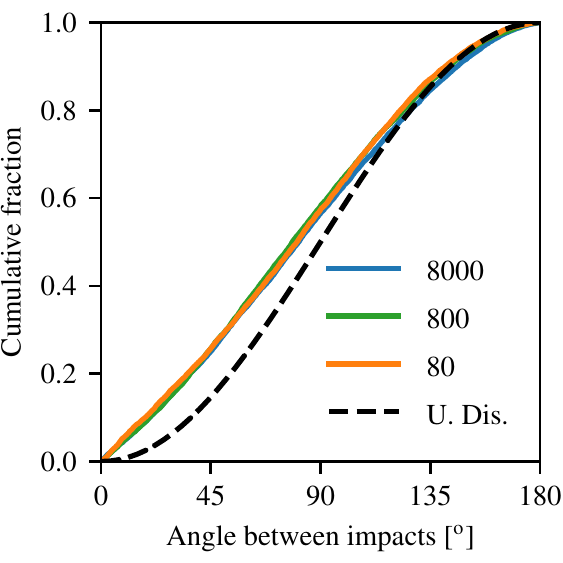}
	\includegraphics{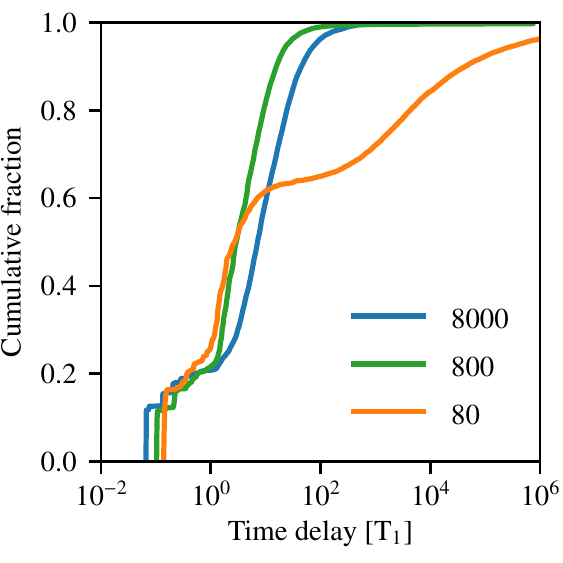}
	\caption{Cumulative distributions of impact angles (\textit{left panels}), offset angle between successive collisions (\textit{center panels}) and time delays between successive collisions (\textit{right panels}) for dynamical evolution sets have to have similar relationships between apocenter-to-Hill radius and radii-to-Hill radius according to the simple scaling relation (see text). The labels provide the value of $\rcollH^{-1}=r_\mathrm{H}/r_\mathrm{coll}$, the dashed black lines labelled "U. Dis." stand for the expected behavior in case of events uniformly distributed in space.}
	\label{fig:map}
\end{figure*}

We further check whether other properties follow the same scaling relation with Figure~\ref{fig:map}. Each row contains three dynamical evolution sets that have the same value of $k$. The upper row shows the sets with $\rapoH=r_\mathrm{apo}/r_\mathrm{H}=0.3$ and $\rcollH^{-1}=8000$ (blue), $\rapoH=0.5$ and $\rcollH^{-1}=800$ (green), and $\rapoH=0.7$ and $\rcollH^{-1}=80$ (orange) while the lower row shows $\rapoH=0.5$ and $\rcollH^{-1}=8000$ (blue), $\rapoH=0.7$ and $\rcollH^{-1}=800$ (green), and $\rapoH=0.9$ and $\rcollH^{-1}=80$ (orange). We can observe the same behavior for the other quantities, with the  distinction that was discussed in the previous paragraph for the case with $\rapoH=0.3$ (the blue curve on the top row), and a change for the extended time delays in the series with $\rapoH=0.9$ (the orange curve in the bottom row). The time delays are slightly shifted from one set to the other due to the change in $\rapoH$ of each set, in the same way as the onsets of each curve are shifted in in Figure~\ref{fig:time-series}, but unlike that result, they otherwise retain the same general shape. The exception being the largest apocenter, given in terms of the Hill radius, which comparatively has more events where the remnants gets far away from eacher other, and returning more like a HRC.
The scaling relation therefore provides a good approximation to determine the overall behavior of GMCs for $\rapoH$ between roughly $0.4$ and $0.8$. Lower apocenters, as shown by the blue curve in the top row, behave more independently of the Hill-to-body radius ratio, since the quantities are closer to the outcome of the prior event.

Based on the results of this section, we may suppose that the underlying dependency on $\rcollH$ is on the orbit eccentricity, which is parameterized by the factor $\rperiH=r_\mathrm{peri}/r_\mathrm{H}$ rather than $\rcollH=r_\mathrm{coll}/r_\mathrm{H}$. The two are closely linked together since $r_\mathrm{peri}\sim r_\mathrm{coll}$. The impact angle, that we chose to be fixed at \ang{60}, along with $\rapoH$ determines the relationship between $r_\mathrm{peri}$ and $r_\mathrm{coll}$. Hence, the eccentricity depends on the impact angle; grazing collisions (where the impact angle is close to \ang{90}), will have lower eccentricity as $r_\mathrm{peri}\approx r_\mathrm{coll}$. Therefore the results shown here also slightly depend on the choice of the impact angle. We expect that for grazing impact angles, the behavior will be shifted as if $\rcollH$ was increased, while we expect the opposite for more head-on impact angles.

\section{Discussion and Conclusion}

In this work, we model the intermediate stage of GMCs, where two bodies orbit each other on a eccentric orbit. Without external perturbation, these two objects collide again after one revolution. If the same bodies orbit around another central body (a star in the case of planets, or a planet in the case of satellites), then the magnitude of the perturbation depends on two quantities: the separation that would be obtained in an isolated case given in terms of the Hill radius (the scaled apocenter distance $\rapoH=r_\mathrm{apo}/r_\mathrm{H}$), and to a lower extent, the size of the Hill sphere compared to the bodies' physical size.

\subsection{Summary of results}

In the simple case when only the central body is present, the following rules can be applied to determine its effect on a GMC.

If the two bodies remain within $1/3$ of the Hill sphere during the revolution ($\rapoH<1/3$), then the orbit is barely modified from the expected outcome modeled assuming a two-body problem only. Modeling such collisions neglecting the presence of other bodies provides sufficient fidelity. This is potentially less correct for instances where the ratio between the Hill radius and the body radii is larger than the order of 1000.

If the scaled apocenter distance is between $1/3$ and $3/4$, then the central body starts to have an effect on the orbit, though not sufficiently to unbind the remnants. As the scaled apocenter distance increases, the impact angle tends toward a uniform distribution and the symmetry plane of the second is shifted with respect to the one of the first encounter. To scale the results obtained in Section~\ref{sec:res} with any particular situation, the relationship from Equation~(\ref{eq:scaling}) can be used to determine the relevant value of $\rapoH$, bearing in mind that results in Section~\ref{sec:res} were obtained with $\rcollH^{-1}\simeq153$. At the upper boundary of this range ($\rapoH\approx3/4$), the impact angle distribution relates much more to what is expected from a uniform distribution than the outcome from the first encounter, and the median of the angle between the collision planes is near \ang{45}.

Beyond $3/4$ of the Hill radius, only a minority of the realisations results in the subsequent collision occurring after a single orbit, and some of the pairs get unbound. Relationship between the outcome of the first encounter and the properties of the subsequent collision no longer holds; assuming uncorrelated events for the impact angle and orientation provides a better fidelity.

\subsection{Prevalence of perturbed GMCs}

We provided in Section~\ref{sec:prev} the relationship between the relative velocity and the distance reached during the intermediate orbit, given in terms of the Hill radius. We discussed that bodies close to the central object have a higher likelihood to have GMCs reaching a sizable fraction of the Hill radius. GMCs are affected when the apocenter is higher than roughly $1/3$ of the Hill radius. Therefore, GMCs are more likely to be affected when they occur close to the central body. The range of velocities is slightly expanded from the discussion in Section~\ref{sec:prev} though. Assuming the same relationship with $r_\mathrm{coll}\approx1.6r_1$, then the range resulting velocity fulfilling the condition $\rapoH>1/3$ is $0.79\lesssim v_\mathrm{coll}/v_\mathrm{esc}<1$ for Ganymede, $0.88\lesssim v_\mathrm{coll}/v_\mathrm{esc}<1$ for Titan, $0.99\lesssim v_\mathrm{coll}/v_\mathrm{esc}<1$ for Earth, and $1-\num{3.5e-4}\lesssim v_\mathrm{coll}/v_\mathrm{esc}<1$ for Pluto. This values means that nearly all GMCs occurring on Ganymede would be affect by Jupiter, while the probability of a perturbed GMC on Pluto is very small.

On the other hand, when the body radii and the Hill radius are less dissimilar, another competing effect takes place. A lower orbital eccentricity is required to attain distances comparable to the Hill radius, and in doing so, the perturbation of the orbit is comparatively lower for a given apocenter given in terms of the Hill sphere, as we obtained in Section~\ref{sec:other}. For instance, in the case where where $\rcollH=r_\mathrm{coll}/r_\mathrm{H}=0.1$ (the lower boundary on Figure~\ref{eq:scaling}), then perturbed GMCs require an apocenter on the order of half the Hill radius. This corresponds to $r_\mathrm{apo}/r_\mathrm{coll}\approx 5$, hence to velocities of $v_\mathrm{coll}/v_\mathrm{esc}\gtrsim0.9$. For comparison, if this correction was left out, the perturbation would start when $r_\mathrm{apo}/r_\mathrm{coll}\approx 3.5$, corresponding to $v_\mathrm{coll}/v_\mathrm{esc}\gtrsim0.84$, leaving a wider range of lower impact velocities for perturbed GMCs.

Finally, in situations where the Hill sphere is on the same order as the radius, or even smaller for instance in case of Saturn's inner moonlets \citep[e.g.,][]{2018NatAsLeleu}, GMC cannot happen. In this situation, there is direct transition between merger and HRC. The remnants may of course collide again, but they would occur under the characteristics of HRC regime, i.e. with an unconstrained geometry, tough the velocity will be related the end state, that is about the mutual escape velocity.

\subsection{Application to specific scenarios}

We return to the specific example provided in Section~\ref{sec:gmc}, namely a target mass $m_\mathrm{tar}=\SI{0.9}{M_\oplus}$, a projectile mass $m_\mathrm{proj}=\SI{0.2}{M_\oplus}$, a impact velocity $v_\mathrm{coll}/v_\mathrm{esc}=1.1$ and an impact angle $\theta_\mathrm{coll}=\ang{52.5}$. In this case we obtain that the intermediate orbit reaches $1/3$ of the Hill sphere if the bodies are located at \SI{1.15}{AU}, assuming a Sun-mass central body. If such collision were to occur at \SI{1}{AU}, it would result in $\rapoH\simeq0.39$. From the results obtained in this work, this would imply that there is roughly 80\% probability for the bodies to miss after a single revolution, although they would likely collide again with a year. After this time, the impact angle of the second collision would likely be between \ang{40} and \ang{90}, with a median close to the values shown in Figure~\ref{fig:gmc-params}. If the collision were to occur even closer to the Sun, the intermediate orbit should be even more perturbed.

Of the GMC scenarios discussed so far, the one that is most significantly affected by the presence of a central body is the scenario of \citep{2013IcarusAsphaug} for the formation of Saturn's middle-sized moons by the coalescence of Titan from two or more major satellites. The authors modeled the relevant collisions in the inertial frame, an assumption that we find is not valid. The Hill radius around Titan, in its present orbit, is  20 Titan radii, while the typical apocenter of the grazing collisions that \citet{2013IcarusAsphaug} model go out to about half that distance, implying strong perturbations by Saturn. The specific collisions that they model are not, in fact, accretionary, but should be counted as hit and run.

The first effect of taking Saturn into account is to reduce the allowable impact angle that can result in a satellite-satellite merger. This would have the tendency of reducing the ejection of middle-sized clumps in the merger that produced Titan, because the kinds of clumpy spiral arms that would produce middle-sized moons are favorably produced during graze and merge events. The parameter space for GMC is reduced by the presence of Saturn, so that the parameter space for creating middle-sized moons in a single giant impact is also reduced. On the other hand, it implies that most giant impacts around Saturn would happen in the form of collision chains \citep{2019ApJEmsenhuberA}, one hit and run collision leading to another collision, until the conditions for merger (low velocity, low impact angle) are met. This could have the effect of enhancing middle-sized-moon production, by setting the system up for multiple giant impacts in a row. In any event this hypothesis of Saturn system and Titan formation must be revisited with the physics of 3-body interactions in mind.

\subsection{Transition between GMC and HRC}

We find that the transition between GMC and HRC is smooth; as the orbit ranges closer to Hill sphere, the probability for a event to get unbound increases. The same smooth transition is obtained other impact properties, such as the impact angle of the second collision and the alignment of the successive encounters. A simple criterion to distinguish GMC and HRC based on the overall accretion efficiency \citep[e.g.,][]{2010ApJKokubo} therefore does not reflect what is occurring while other bodies are present.

From an accretion point of view, the transition between GMC and HRC is not defined by whether the remnants are bound or not, but rather by the distance those would range if they were isolated, compared to the Hill radius of the actual situation. If these two values are equal, then the most likely outcome is that the bodies will get unbound at some point during the orbit, hence resulting in an HRC.

From the point of view of \textit{N}-body studies, GMCs whose remants reach a substantial fraction of the Hill radius should not be treated as unconditional mergers. The best would be to treat them as HRCs, by putting the remnants of the first encounter back in the \textit{N}-body. This ensures that the transition between GMC and HRC is properly represented, and has the further advantage to capture the same physics in both regimes. The downside is that this assumes that the different phases of a GMC (collisions themselves and the intermediate orbit) can modeled separately; this might not be entirely correct when Hill radius is not much larger than the body sizes.

\acknowledgments

The authors thank Saverio Cambioni and Adrien Leleu for fruitful discussions, and an anonymous reviewer for the helpful comments and edits that improved the manuscript.
We acknowledge support from NASA grant NNX16AI31G and the University of Arizona.
An allocation of computer time from the UA Research Computing High Performance Computing (HPC) is gratefully acknowledged.

\software{Mercury \citep{1999MNRASChambers}, matplotlib \citep{2007CSEHunter}}

\bibliographystyle{aasjournal}
\bibliography{return}

\appendix

\section{Searching for bound bodies}
\label{sec:mix}

We presented in \citet{2019ApJEmsenhuberA} a method to reliably distinguish Smoothed Particle Hydrodynamics (SPH) bodies that a barely unbound, by preceding the gravity search by friends-of-friends (FoF) search, so that bodies are not split between bound and unbound components due to their spin. In graze and merge collisions (GMCs), we face a different, but related difficulty: the main remnants are bound, so the gravity search returns one single massive clump that encompasses both. The FoF search on the other hand can distinguish those, yet it leaves particles with too low density unaccounted for.

So to discriminate the two remnants in a GMC without underestimating their mass, we adopt the following procedure. We first perform FoF and gravity search as previously, and we a third step to combine both results. The algorithm is this new step is as follows: for every gravitational clump a list of FoF clumps is made. If no FoF clump is found, then nothing is done. If a single FoF clump is found, then all SPH particles in the corresponding gravity clump are added to it, and the bulk density is kept the same as in the FoF clump. If multiple FoF clumps are found, then particles gravitationally bound together, but that are not part of any FoF are added to the FoF with whom they have the lowest total energy. In this case the bulk density is taken from the FoF clump, which is needed to infer the radius for the N-body modeling.

\begin{figure*}
	\centering
	\includegraphics{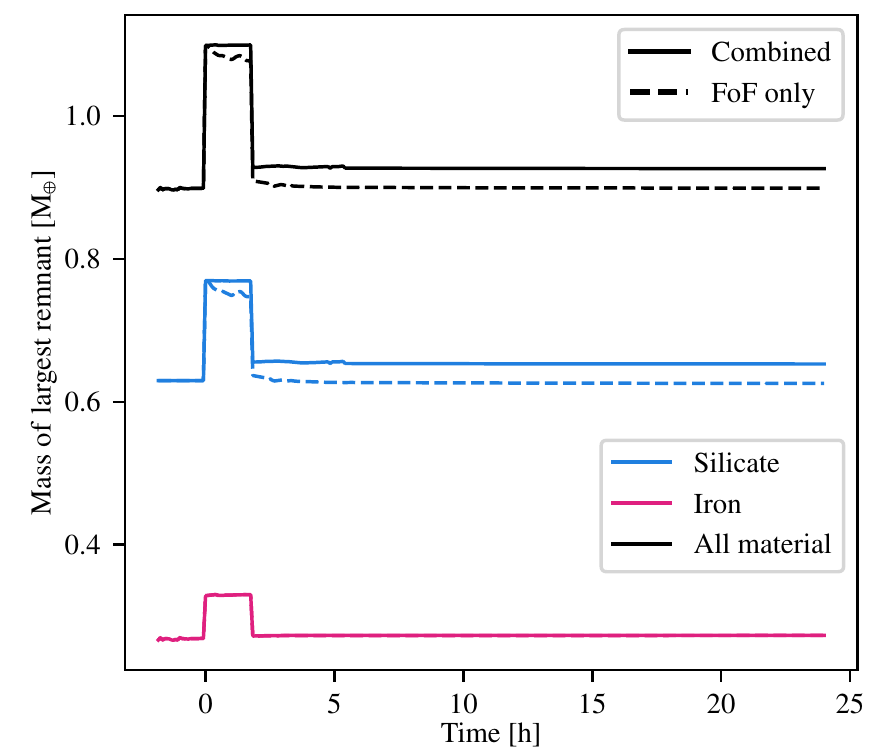}
	\includegraphics{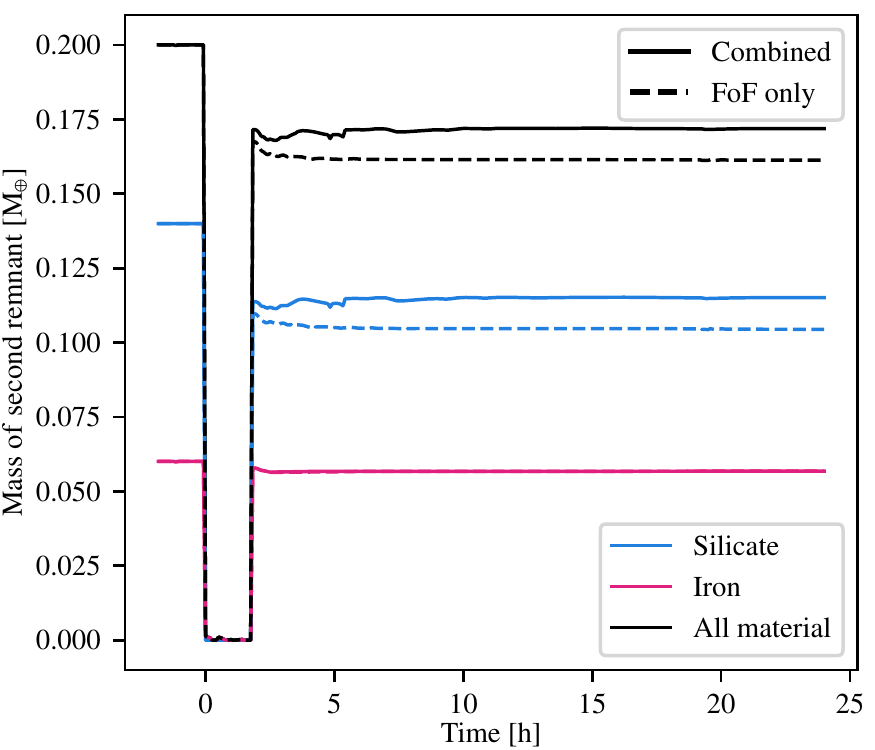}
	\caption{Comparison of the masses for the largest body (\textit{left panel}) and second largest (\textit{right panel}) for one graze and merge collision modeled using Smooted Particle Hydrodynamics (SPH) using two different methods (see text): simple friends-of-friends (FoF) search (dashed lines) and recombination with gravity search (solid lines).}
	\label{fig:mix}
\end{figure*}

The difference between this approach and using only a FoF search is shown in Figure~\ref{fig:mix}. There is almost no differences during the pre-collision stage (negative times), as the initial bodies are well recovered. However, after the collision, using only a FoF search misses several percents of the overall mass, and accounting for bound material that is not found by that search is needed to avoid artificially low body masses in GMC during the intermediate stage, i.e. between the initial and return impacts. We also note that the material that is neglected is mostly the upper mantle, and therefore neglecting the remaining material may lead to incorrect composition determinations.

\end{document}